\documentclass{aa}  
\usepackage{graphicx}
\usepackage{txfonts}
\usepackage{subcaption}
\usepackage{verbatim}

\usepackage[switch]{lineno}
\usepackage{datetime}
\usepackage{setspace}
\usepackage[normalem]{ulem}
\usepackage{soul}
\usepackage{hyperref}
\usepackage{multirow}

\usepackage{orcidlink}

\usepackage[usenames,table,dvipsnames]{xcolor}
\definecolor{blue}{RGB}{66, 153, 233}
\definecolor{red}{RGB}{255, 0, 0}
\definecolor{purple}{RGB}{255, 0, 255}

\newcount\Comments  
\newcommand{\kibitz}[2]{\ifnum\Comments=1\textcolor{#1}{#2}\fi}

\newcolumntype{P}[1]{>{\centering\arraybackslash}p{#1\textwidth}}

\begin{document} 
    \title{The XRISM view of the front and rear ejecta dynamics\\ in Tycho’s SNR}

   \author{L. Godinaud \inst{1, 2}\,\orcidlink{0009-0005-0005-0272},
   A. Decourchelle \inst{1}\,\orcidlink{0000-0002-1796-758X},
   R. Giuffrida \inst{1}\,\orcidlink{0000-0002-2774-3491},
   T. Tanaka \inst{3}\,\orcidlink{0000-0002-4383-0368},
   C. Badenes \inst{4}\,\orcidlink{0000-0003-3494-343X},
   G. Ferrand \inst{5,6}\,\orcidlink{0000-0002-4231-8717},
   J. Ballet \inst{1}\,\orcidlink{0000-0002-8784-2977},
   T. Holland-Ashford \inst{7}\,\orcidlink{0000-0002-7643-0504},
   A. Bamba \inst{8}\,\orcidlink{0000-0003-0890-4920},
   B. J. Williams \inst{7}\,\orcidlink{0000-0003-2063-381X},
   T. Mizuno \inst{9}\,\orcidlink{0000-0001-7263-0296},
   F. Acero \inst{1, 10}\,\orcidlink{0000-0002-6606-2816},
   J. Vink \inst{11, 12}\,\orcidlink{0000-0002-4708-4219}}

    \institute{
Université Paris-Saclay, Université Paris Cité, CEA, CNRS, AIM, F-91191 Gif sur Yvette, France
    \and
Smithsonian Astrophysical Observatory, Center for Astrophysics | Harvard \& Smithsonian,  MS-3, 60 Garden St., Cambridge, MA 02138, USA
    \and
Department of Physics, Konan University, Hyogo 658-8501, Japan
    \and
Department of Physics and Astronomy and Pittsburgh Particle Physics, Astrophysics and Cosmology Center (PITT PACC), University
of Pittsburgh, 3941 O’Hara Street, Pittsburgh, PA 15260, USA
    \and
Department of Physics and Astronomy, 30A Sifton Road, University of Manitoba, Winnipeg, Manitoba, R3T 2N2, Canada
    \and
RIKEN Center for Interdisciplinary Theoretical and Mathematical Sciences (iTHEMS), 2-1 Hirosawa, Wako, Saitama 351-0198, Japan
    \and
Astrophysics Science Division, NASA Goddard Space Flight Center, Greenbelt, MD 20771, USA
    \and
Department of Physics, University of Tokyo, Tokyo 113-0033, Japan
    \and
Hiroshima Astrophysical Science Center, Hiroshima University, Hiroshima 739-8526, Japan
    \and
FSLAC IRL 2009, CNRS/IAC, La Laguna, Tenerife, Spain
    \and
Anton Pannekoek Institute/GRAPPA, University of Amsterdam, Science Park 904, 1098 XH Amsterdam,  The Netherlands
    \and
SRON Netherlands Institute for Space Research, Niels Bohrweg 4, 2333 CA Leiden, The Netherlands
    }

   \date{Submitted 18 September 2026; accepted 18 November 2026 }

  \abstract
{We present a detailed line-of-sight dynamics study of Tycho's supernova remnant (SNR) with \emph{XRISM/Resolve} instrument, based on five different pointings obtained during the first two years of the mission. Thanks to a spectral resolution better than 5 eV, \emph{Resolve} spectra reveal a clear Doppler splitting of the Si, S, Ar, Ca and Fe emission lines towards the center of the remnant, allowing us to study independently the front and back sides of the shocked ejecta for the first time. Using a phenomenological Gaussian line model and a Bayesian fitting method, we analyze the broadband high resolution spectra for each pointing and for the 149 individual pixels.

We derive the dynamics and flux of the front and back sides of the shocked ejecta independently for each element, as well as their global line broadening.
The dynamics of Tycho's SNR present asymmetries, with the back side expanding faster than the front side for Si, S and Fe. We reveal a line-of-sight stratification of the elements with Fe at lower velocities than Si and S. A strong front/back flux asymmetry is observed for Si, with the front side being brighter than the back side, by up to a factor two in the northeastern part of the remnant. This flux asymmetry is also found for S albeit with weaker contrasts. For Fe, we observe spatial variations in the flux ratio, with either the front or the rear side dominating. 

Measuring the properties of the front and back sides using \emph{XRISM/Resolve} allows reassessing the interpretation of the average kinematics along the line of sight obtained at CCD resolution, which is affected by the flux asymmetry between the front and rear sides and not only a velocity asymmetry.
We combine our line-of-sight velocity measurements from \emph{XRISM/Resolve} with proper motion measurements from \emph{Chandra} to reconstruct the three-dimensional dynamics of Tycho's supernova remnant.
Finally, we discuss how these new results set constraints on the explosion model and the progenitor system at the origin of Tycho’s SNR.
}


   \titlerunning{}
   \authorrunning{Godinaud et al.}
   \maketitle
%

\section{Introduction}

Type Ia supernovae (SNe Ia) and core-collapse supernovae (CCSNe) represent the two main categories of stellar explosions. While CC-SNe arise from the gravitational collapse of massive stars, SNe Ia are thermonuclear events. The subclass called "normal" SNIa are used as standard candles for measuring cosmic distances. 
Two progenitor scenarios are typically considered to explain SNe Ia: the single-degenerate (SD) channel, in which a white dwarf accretes matter from a non-degenerate companion star, and the double-degenerate (DD) channel, in which two white dwarfs interact \citep[for a review see][]{Liu2023}. 
In both SD and DD models, the mass loss history of the progenitor system has the potential to modify the structure of the CSM, even though there is a wide range of predictions, ranging from a relatively undisturbed ambient medium in DD models whose final orbital decay is mediated by gravitational wave emission to dense structures created by mass transfer or common envelope ejection right before the explosion in SD models. However, a CSM may also be present in some DD scenarios in the case of SNe Ia-CSM subclass \citep[e.g. the study of SN 2020aeuh by][]{Tsalapatas2025}.

As with the progenitor system, the explosion mechanism of Type Ia supernovae also remains an open question.
The first explosion mechanisms studied were pure detonations in near-Chandrasekhar mass white dwarfs \citep{Arnett1969}. However, the resulting yields did not match observations \citep{Arnett1969}. Pure-deflagration models were later developed \citep{Nomoto1984}, but their properties do not appear to correspond to those of normal Type Ia supernovae.
More recently, delayed-detonation models \citep{Khokhlov1991} have proven very promising in explaining the physics of these objects. Explosions of sub-Chandrasekhar mass white dwarfs through a double-detonation mechanism \citep{Sim2010, Kromer2010} are also competitive and agree well with current observations.
A complete review is presented in \cite{Ruiter2025}.
These different models directly impact the element yields produced by nucleosynthesis, which are often used to constrain explosion models.
Furthermore, the spatial distribution of chemical elements strongly depends on the explosion mechanism. Pure deflagration models predict little to no element stratification \citep{Fink2014}, whereas delayed detonation models \citep{Seitenzahl2013} produce an inner Fe-rich core surrounded by layers of intermediate-mass elements (IMEs). Double-detonation models predict an even stronger stratification \citep{Tanikawa2018}, with a Fe core, an IME layer stratified from Ca to Si, and an outer Ca-rich shell. 

The study of supernova remnants (SNRs), and especially the young ones (a few hundred years old) is a powerful tool to constrain the progenitor system and explosion mechanism. The expansion of the stellar material ---the ejecta--- drives a forward shock into the surrounding environment, while its interaction with the ambient medium produces a reverse shock that propagates inward into the ejecta. Young SNRs appear as shocked shells of finite thickness. Shocked ejecta and shocked surrounding environment being heated to tens of million degrees, SNRs are bright X-ray sources.

Although Type Ia remnants are generally more spherical than those from CC-SNe in X-rays \citep{Lopez2011}, they nonetheless can exhibit asymmetries that contain important diagnostic information as shown in three-dimensional numerical simulations of SNRs \citep{Ferrand2019, Ferrand2021, Ferrand2022}. These asymmetries manifest in flux, element distribution, and dynamics, but are challenging to disentangle in observations because of the flux-weighted integration of the 3D information in the line of sight.
Morphological and dynamical asymmetries can also be linked to interactions with an inhomogeneous circumstellar medium (CSM), which may itself bear the imprint of the progenitor system \citep{Chiotellis2013, Court2025}. 
In addition, some models predict a bulk velocity of the ejecta in the range of 30 to more than 1000 km s$^{-1}$ \citep{Shen2018}. This could induce velocity asymmetries even in a homogeneous medium.

In this work we focus on Tycho’s SNR, a prototypical "normal" Type Ia remnant \citep{Krause2008, Rest2008, Badenes2006} whose explosion was observed in 1572 by Danish astronomer Tycho Brahe. X-ray observations reveal strong emission lines from Si, S, Ar, Ca, and Fe as observed with \emph{ASCA} \citep{Hwang1997}, \emph{XMM-Newton} \citep{Decourchelle2001}, \emph{Chandra} \citep{Hwang2002} and \emph{Suzaku} \citep{Tamagawa2009}. The Fe distribution is more centrally concentrated, indicating a chemical stratification between the heavier elements and the IMEs \citep{Decourchelle2001, Hayato2010}.
The global spectral X-ray properties of Tycho’s SNR have been used to constrain different explosion scenarios \citep{Badenes2006}.
The abundances of the various observed elements have been compared with the yields predicted by different nucleosynthesis models \citep{Holland-Ashford2025}. However, the explosion mechanism is still under discussion, particularly between delayed-detonation models and double-detonation models, as well as the mass of the white dwarf (Chandrasekhar versus sub-Chandrasekhar). The nature of the progenitor system (SD versus DD) also remains an open question : the absence of an observed surviving companion star \citep{Kerzendorf2018} favors a DD scenario while the presence of a surrounding medium that appears to be shaped by a complex CSM \citep{Williams2016} likely supports an SD scenario.

In addition to these measurements, the study of the 3D dynamics of Tycho’s SNR provides valuable information on both large and small scales.
In the plane of the sky, the proper motion $V_{xy}$ of the forward shock \citep{Katsuda2010, Williams2016}, using \emph{Chandra} observations at different epochs, presents an asymmetry between the eastern and western rims correlated with gradients in the ambient density \citep{Williams2013, Williams2016}.
Moreover, at the northeastern edge, dynamical evidence exists for ejecta interaction with an overdense ambient medium \citep{Godinaud2023}, also detected in radio \citep{Zhou2016, Arias2019, Bo2026}. 
Along the line of sight, the velocity can be measured via the Doppler shift. It is important to note that what we actually measure is a line-of-sight flux-weighted mean velocity. At CCD spectral resolution, this can lead to averaging the properties of the front and back sides if the spectral resolution does not allow them to be distinguished.
\emph{Chandra} Doppler velocity measurements revealed a large-scale asymmetry, with the northern side of the remnant being predominantly blueshifted and the southern side redshifted, whose origin is still open \citep{Godinaud2023, Godinaud2024}.

The XRISM telescope \citep[JAXA, NASA, with ESA contributions; launched in 2023; ][]{Tashiro2025} is revolutionizing the study of X-ray sources in the 2-15 keV domain. It is equipped with a microcalorimeter detector, \emph{Resolve} \citep{Ishisaki2022, Porter2024}, which achieves a spectral resolution of 5 eV.
During the Performance and Verification phase, several SNRs were observed, revealing new information about their thermal properties and 3D dynamics. Among these remnants are N132D \citep[First light,][]{XRISMCollaboration2024_N132D}, Cas A \citep{Vink2025, Suzuki2025, Bamba2025, Plucinsky2025}, SN1987A \citep{XrismCollaboration2025_SN87a} and W49B \citep{XrismCollaboration2025_W49B}.
\emph{XRISM/Resolve} observations of Tycho’s SNR clearly separated the blueshifted and redshifted Fe-K emission towards the remnant’s center (XRISM collaboration et al, 2026, in prep.). This line is split by the Doppler effect, enabling independent mapping of the properties of the front and back sides of the SNR. The first XRISM paper on Tycho’s SNR focused on the Fe line using the Performance and Verification observations.
The present study uses the Performance and Verification and AO1 observations of Tycho's SNR to conduct a spatially resolved study of the front and back shocked ejecta morphology and dynamics per element and in particular for the Si, S and Fe. The scientific objectives are the following: characterizing  the three-dimensional flux and velocity asymmetries, and investigating the degree of stratification between IMEs and Fe.

\section{Data, reduction, and fitting tool}

\begin{table*}[h!]
\caption{\footnotesize List of the XRISM observations used in this study and their properties.}
\label{tab:ListPointing}
\centering
\begin{tabular}{cccccccc}
\hline
Name & Cycle & ID & Start date & Exposure & RA (deg) & Dec (deg) & Angle (deg)\\
 & &  &  & time (ks) &  &  &  \\
\hline
NorthWest & PV & 000160000  & 2024 February 2  & 204 & 6.233 & 64.168 & 133.6  \\
Center & PV & 300060010 & 2024 March 2 &  175 & 6.334 & 64.138 & 140.6 \\
North & PV & 300062010 & 2024 March 22 & 42 & 6.334 & 64.191 & 175.1 \\
South & PV & 300063010 & 2024 March 23 & 46 & 6.334 & 64.087 & 176.3 \\
\hline
NorthEast & AO1 & 201118010  & 2024 September 23  & 120  & 6.408 & 64.171 & -4.9 \\
\hline
\end{tabular}
\vspace{-0.2cm}
\end{table*}

\begin{figure}[h]
\centering
\includegraphics[scale=0.28, trim = 0 0 0 0, clip=true]{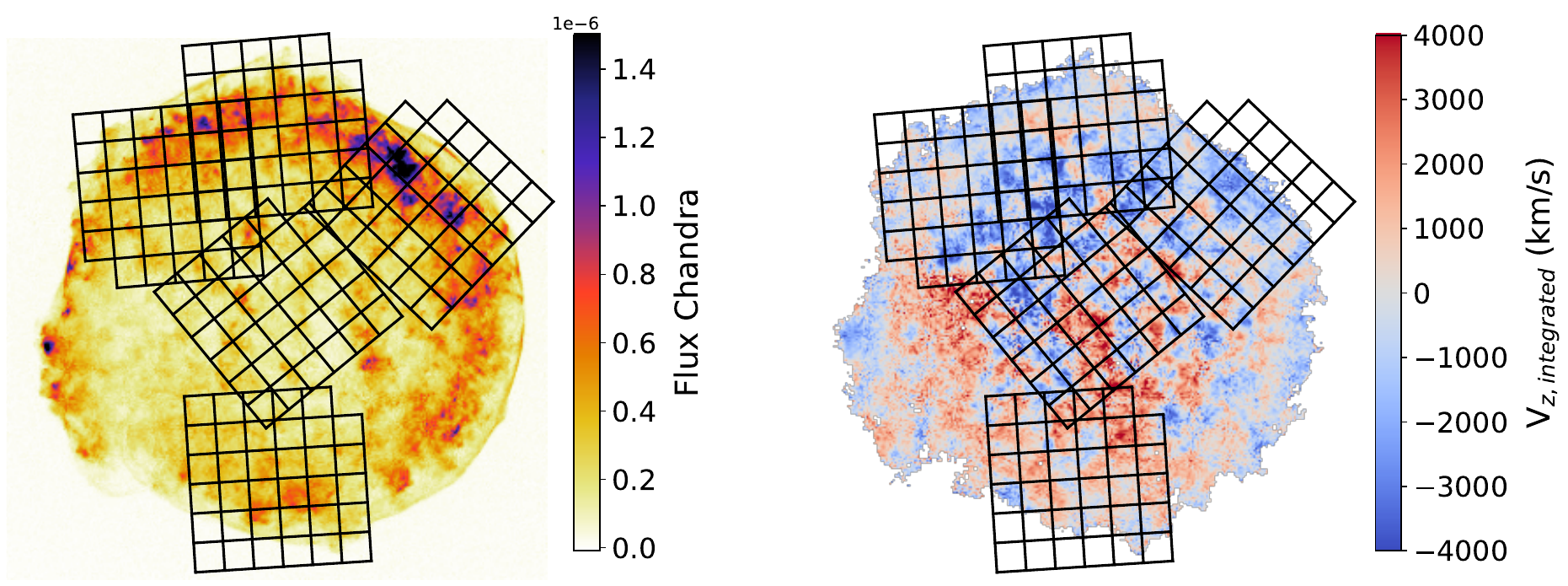}
\caption{\footnotesize \emph{Left:} \emph{XRISM/Resolve} pointings overlaid on the \emph{Chandra} flux map between 0.5 and 7 keV. \emph{Right:} \emph{XRISM/Resolve} pointings overlaid on the line-of-sight mean Si velocity \emph{Chandra} map \citep{Godinaud2023}.}
\label{fig:PointingXRISM}
\vspace{-0.2cm}
\end{figure}

In this paper, we use \emph{XRISM/Resolve} data from the four pointings performed during the Performance and Verification phase (Center, NorthWest, North, South), plus an additional pointing obtained in AO1 at the NorthEast (PI. A. Decourchelle). The properties of these observations are summarized in Table \ref{tab:ListPointing}. 
Figure \ref{fig:PointingXRISM} shows \emph{XRISM/Resolve} pointings overlaid on \emph{Chandra} 0.5-7 keV image (left panel) of Tycho’s SNR. 65\% of the remnant is covered by the field of view of these pointings. The right panel of Figure \ref{fig:PointingXRISM} shows these five pointings overlaid on the line-of-sight mean Si velocity map obtained by \cite{Godinaud2023}.

The data were reprocessed using the HEASoft 6.34 software with calibration from XRISM CalDB 9. Only high-resolution primary grade events were retained. Redistribution matrix files (RMFs) were generated in large format, and ancillary response files (ARFs) were produced using the \emph{Chandra} broadband (1.75–7 keV) image of Tycho’s SNR.
The non X-ray background is not considered here, as its level is significantly lower than the signal of interest. Any background lines that could potentially contaminate the analysis are very narrow, and we have verified that their positions do not affect our results.
At each stage of this analysis, we exclude pixels 12 (calibration pixel) and 27 (unstable pixel) when extracting the spectra.

To perform the spectral fitting, we use the nested sampling algorithm as implemented in the Bayesian X-ray Analysis tool \citep[BXA,][]{Buchner2014} in combination with Sherpa models \citep{Siemiginowska2024} which provide a wrapper to the Xspec models. This tool prevents convergence toward local minima and provides a global view of the statistical landscape, thereby enabling a robust estimation of uncertainties. We use priors on each parameter that are sufficiently broad to avoid biasing our results. For the redshifts, the priors are uniform, and for the normalizations, they are log-uniform.
BXA also allows us to identify degeneracies in the model. However, BXA is computationally expensive and requires the use of computing clusters: in this study, one spectral fitting takes around 14 hours with a parallelization on 32 CPUs.

In the following, errors are reported at a 68.3\% confidence level (1$\sigma$), unless otherwise specified.

\section{Methodology}
\label{section:Methodo}

\subsection{The Gaussian spectral models}
\label{subsection:GaussianModels}

\begin{figure*}
\sidecaption
\includegraphics[width=12cm, scale=1.5, trim = 45 5 90 15, clip=true]{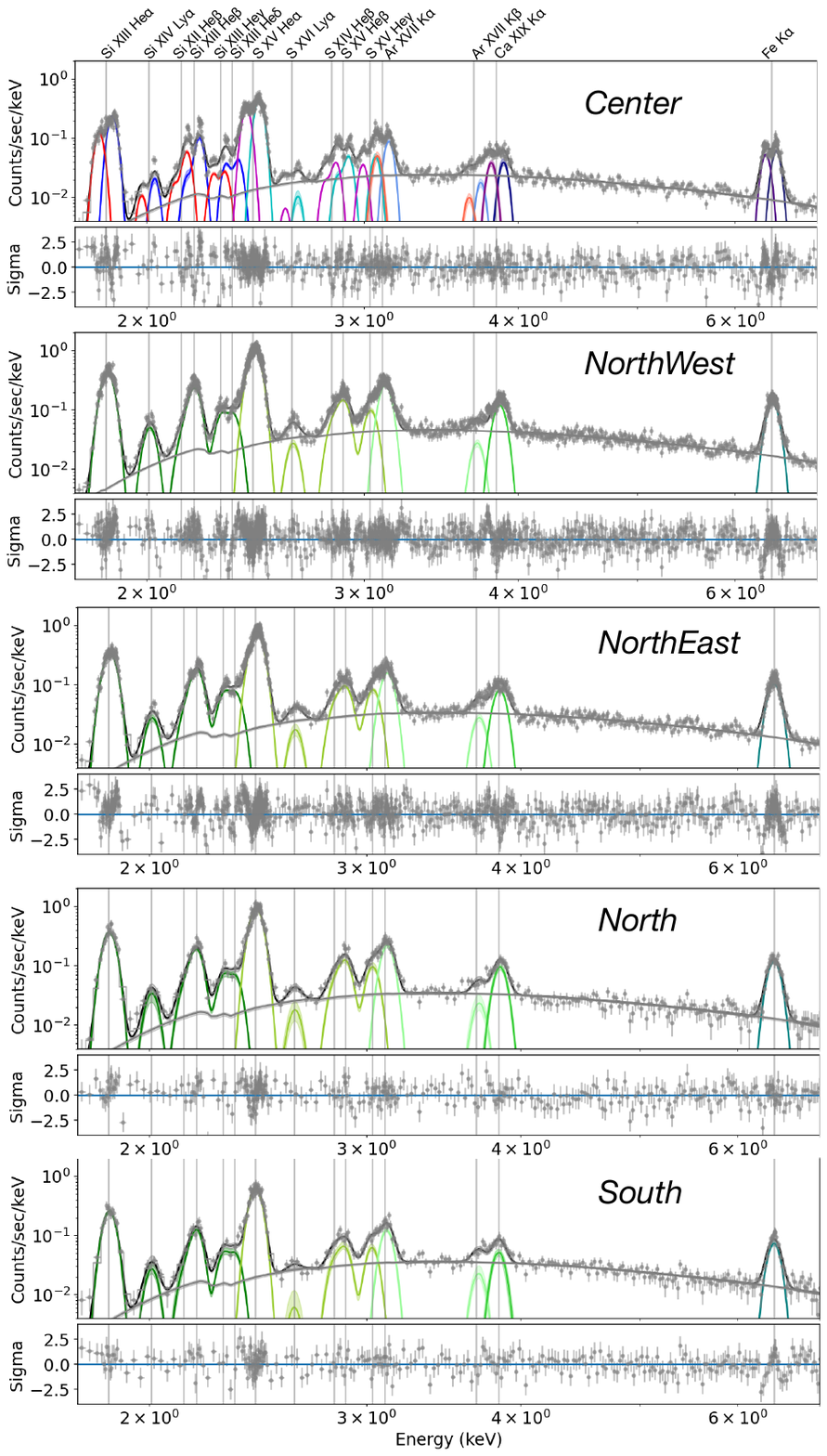}
\caption{\footnotesize 
Best-fit of the overall spectra for the Center, NorthWest, NorthEast, North, and South pointings (from top to bottom). Gray vertical lines indicate each emission line at its rest energy, modeled with redshifted (in red/reddish) and blueshifted (in blue/bluish) Doppler components for the center pointings and a single Doppler shift component (in green) for all the other pointings. The variations within green, red and blue colors indicate emission lines from different elements. Each Gaussian has its associated uncertainty envelope at 90\% confidence level. The energy axis is logarithmic.}
\label{fig:FirstViewSpectra}
\end{figure*}

The global spectra for the Center, NorthWest, NorthEast, North and South pointings are presented in Figure \ref{fig:FirstViewSpectra}. At the \emph{XRISM/Resolve} spectral resolution, a multitude of spectral lines appears, previously inaccessible with the CCD resolution of \emph{Chandra} and \emph{XMM-Newton}.  Towards the center of Tycho’s SNR, the Si, S, Ar, Ca and Fe K-lines appear split in blueshifted and redshifted components. These Doppler components  correspond respectively to the front and rear sides of the SNR emitting ejecta shell. In contrast, towards the edges, the lines are not observed to be split. There, the line of sight does not pass through two separate front and rear shocked shells, but continuously integrates the shocked shell material.

We conduct a broadband analysis between 1.75 and 7 keV, using a spectral model consisting of a series of Gaussian lines associated to ejecta emission and a continuum. The continuum is dominated by the synchrotron emission of TeV electrons accelerated at the forward shock, and is modeled by a powerlaw, whose spectral index is free \citep{Bamba2005, Tamagawa2009}. 
The emission from the shocked ambient medium, contributing to less than $\sim$5\% of the total emission at 2 keV \citep{Ellien2023, Holland-Ashford2025}, is negligible.
We identified 15 lines in the \emph{XRISM/Resolve} spectra, shown as gray vertical lines in Figure \ref{fig:FirstViewSpectra}. Each Gaussian corresponds to a line complex, rather than to a unique line. For the spectrum of the central pointing, it is necessary to introduce two Gaussian components per emission line in order to reproduce the line splitting observed between the redshifted (rear) and blueshifted (front) sides of the remnant. For the spectra of the edge pointings, only one Gaussian component per emission line is required. 

To get the kinematic properties of the plasma, we measure the line-of-sight Doppler shifts of the Gaussian lines assuming a rest energy for each emission line, whose value depends on the temperature and ionization state of the plasma. To define this value, we used the Python package linked to the atomic database AtomDB as presented in Appendix \ref{Appendix : Eref}. The line identification and rest energies are presented in Table \ref{tab:ListLines}.

\begin{table}[h!]
\caption{\footnotesize Lines fitted in this study, with \emph{XRISM/Resolve} for Tycho's SNR between 1.75 keV and 7 keV.}
\label{tab:ListLines}
\centering
\begin{tabular}{ccccc}
	\hline
	Element & Main ions & Line transitions & E$_{ref}$ (keV) \\
    \hline
    \multirow{6}{2em}{\centering Si}
	  & XIII + XII & He$\alpha$ + DR satellite &1.853 \\
      & XIV & Ly$\alpha$ & 2.006 \\
      & XII & He$\beta$ DR satellite & 2.132 \\
      & XIII & He$\beta$ & 2.183 \\
      & XIII & He$\gamma$ & 2.294 \\
      & XIII & He$\delta$ & 2.346 \\
    \hline
    \multirow{5}{2em}{\centering S}
	  & XV + XIV & He$\alpha$ + DR satellite & 2.439 \\
      & XVI & Ly$\alpha$ & 2.623 \\
      & XIV & He$\beta$ DR satellite & 2.825 \\
      & XV & He$\beta$ & 2.884 \\
      & XV & He$\gamma$ & 3.033 \\
	\hline
    \multirow{2}{2em}{\centering Ar}
	  & XVII & K$\alpha$ & 3.103 \\
      & XVII & K$\beta$ & 3.685 \\
	\hline
	  Ca & XIX  & K$\alpha$ & 3.839  \\
	\hline
	  Fe & XV--XXI & K$\alpha$ & 6.427 \\
	\hline
\end{tabular}
\vspace{-0.2cm}
\end{table}

To reduce the large number of free parameters of the model, we introduced several constraints. We assume that all lines of a same element have the same Doppler shift but fit independently the two (red and blue) Doppler shifts for the center region. For each element, the normalization of the front and rear components is left free. However, all lines of a given element are assumed to share the same front-to-rear normalization ratio, which is equivalent to assuming the same temperature and ionization state for the front and back emission, corresponding to a temperature of 1.2 keV and an ionization time of 2e+10 cm$^{-3}$ s in this study \citep{Yamaguchi2017, Uchida2024, Godinaud2024}. Assuming the same front/back ratio for all lines of a given element will presumably make the fit worse in the secondary lines (like He beta). However, our velocity measurements depend mostly on the main He alpha lines, which these second-order residuals affect little. Finally, we assume a global line broadening to all Gaussian lines. We thus set to zero all individual Gaussian line broadenings, and apply to all lines a unique global \emph{gsmooth}, with index fixed at 1 (meaning that the broadening is proportional to the energy). This is an approximation, as the line broadening depends, among other factors, on the considered line complex (as shown in Figure \ref{fig:AtomDB}), on the element or on different physical properties between the front and rear components. 
By examining the posterior distributions for each fitted spectrum, it appears that the broadening parameter is well constrained and is not correlated with the other fitted parameters, for all elements.

In the end, the two-Gaussian model (used toward the center) and the single-Gaussian model (used near the edge) have 32 and 22 free parameters, respectively, which we fit using the BXA tool.

\subsection{Global and pixel level fitting analysis}
\label{subsection:Application}

First, we apply these Gaussian models to fit the global spectra of each of the five \emph{XRISM/Resolve} pointings to characterize their kinematic properties by taking full advantage of the statistics per pointing. Figure \ref{fig:FirstViewSpectra} shows the resulting fits to the global spectra per pointing. Throughout this paper, the results obtained with the two-component (blueshifted and redshifted) Gaussian model are shown in blue (or bluish) and red (or reddish), respectively. The results from the single-component Gaussian model are shown in green. Table \ref{tab:FitGlobal} provides the best-fit parameters resulting from the modeling of the global spectra of each pointing.
We perform in Appendix \ref{Appendix : SiLyalpha_fit} the same analysis for the Si Ly$\alpha$ line only. Indeed, the rest energy of this line is independent of the temperature and ionization timescale and is only due to the Doppler velocity.
This line thus provides a measure of the Si velocity, independent of possible bias on the assumed rest energy. However, due to limited statistics, it cannot be used at pixel level.

\begin{figure}[t!]
\centering
\includegraphics[scale=0.4, trim = 0 0 0 0, clip=true]{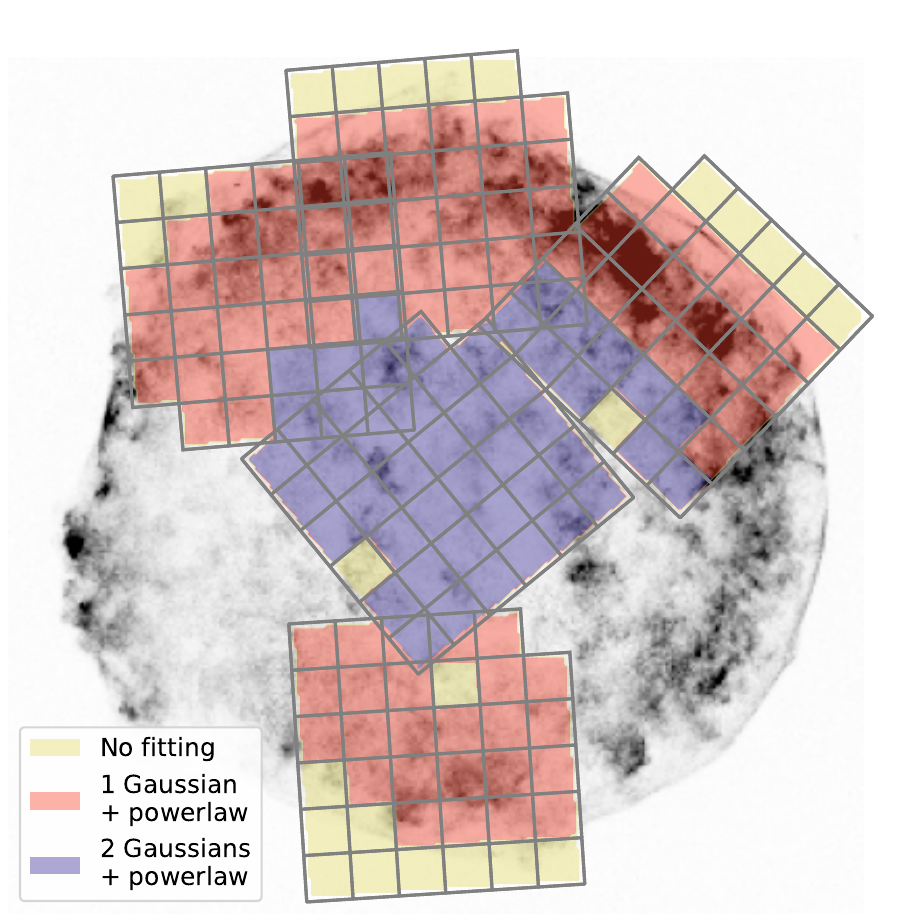}
\caption{\footnotesize Map of the best models used for the pixel-level analysis, superimposed on the \emph{Chandra} 0.5-7 keV image: either no fit is performed \emph{(yellow pixels)}, a single Gaussian for each line and a powerlaw are used \emph{(red pixels)}, or two Gaussians for each line (and a powerlaw) are used to reproduce the Doppler shifts of the front and back sides of the remnant \emph{(purple pixels)}.}
\label{fig:ModelUsed}
\vspace{-0.2cm}
\end{figure}

As the main goal of this study is to map the dynamics of the remnant at the best spatial resolution in order to understand the large and small scales properties of the SNR, we perform our analysis by fitting the \emph{XRISM/Resolve} spectra for each pixel.
Excluding pixels located outside the remnant, this results in a total of 149 high resolution spectra to fit with the models presented in Section \ref{subsection:GaussianModels}.
For pixels located near the SNR's edge, where we expect the line-of-sight emission to encompass a continuous range of shocked material, we do not separate the front/rear components of the emission and a single emission line is indeed observed. 
To address this, we fitted all pixels with the two-Gaussian and the single-Gaussian models, in order to choose the optimal model for each pixel. For regions with an overlap of two pixels from two different pointings, we only present the result obtained from the spectrum with the longest exposure time. In total, we performed around 300 fits at high spectral resolution using BXA.
By examining residuals in the spectra and quality of the posterior distributions for each pixel and the radial profile of the normalization (see Appendix \ref{Appendix : Norme profile}), we define the best model for each pixel, as shown in Figure \ref{fig:ModelUsed}. For the pixels in the southern and northern pointings located near the center, the statistics are insufficient to constrain the two-Gaussian model.

Since Tycho's SNR is currently the extended source with the largest number of XRISM pointings, and because we have enough statistics to carry out this study at the pixel level, this work is likely one of those fitting the highest number of \emph{XRISM/Resolve} spectra.
Automating the process is therefore absolutely necessary. The large number of spectra to be fitted justifies the use of a simplified model with Gaussians as a first step. Indeed, a physical model would be much more complex to converge, especially for the lower statistics in some pixels.

\begin{table*}[h!]
\centering
\caption{\footnotesize Results of the fits on the broadband spectrum for global pointings.}
    {
\begin{tabular}{P{0.07}P{0.15}P{0.12}P{0.12}P{0.12}P{0.12}P{0.12}}
	\hline
	Elt & Parameter & Center & NorthWest & NorthEast & North & South \\
    \hline
    \multirow{5}{1em}{\centering Si} &
    \multirow{2}{3em}{\centering Vz$\ $(km/s)} 
	  & -3320 $\pm$ 60 &\multirow{2}{6em}{\centering -390 $\pm$ 40} & \multirow{2}{6em}{\centering -270 $\pm$ 60 } & \multirow{2}{6em}{\centering 20 $\pm$ 100 }
      & \multirow{2}{6em}{\centering -60 $\pm$ 120 } \\
      & & 3920 $\pm$ 100 &  &  \\
     & \multirow{2}{8em}{\centering Flux$\ $(ph cm$^{-2}$ s$^{-1}$)} 
	  & 5.6e-3 $\pm$ 1.4e-4 &\multirow{2}{7em}{\centering 1.5e-2 $\pm$ 2.3e-4} & \multirow{2}{7em}{\centering 1.4e-2 $\pm$ 2.4e-4 } & \multirow{2}{7em}{\centering 1.4e-2 $\pm$ 4.9e-4 }
      & \multirow{2}{7em}{\centering 9.9e-3 $\pm$ 4.0e-4 } \\
      & & 4.2e-3 $\pm$ 1.4e-4 &  &  \\
      &  Flux ratio & 0.57 $\pm$ 0.01 & - & - & - & - \\
    \hline
    \multirow{5}{1em}{\centering S} &
    \multirow{2}{3em}{\centering Vz$\ $(km/s)} 
	  & -3250 $\pm$ 50 &\multirow{2}{6em}{\centering -520 $\pm$ 30} & \multirow{2}{6em}{\centering -320 $\pm$ 50 } & \multirow{2}{6em}{\centering -60 $\pm$ 80 }
      & \multirow{2}{6em}{\centering 270 $\pm$ 90 } \\
      & & 3830 $\pm$ 80 &  &  \\
    & \multirow{2}{8em}{\centering Flux$\ $(ph cm$^{-2}$ s$^{-1}$)} 
	  & 1.3e-3 $\pm$ 2.2e-5 &\multirow{2}{7em}{\centering 4.2e-3 $\pm$ 3.8e-5} & \multirow{2}{7em}{\centering 3.3e-3 $\pm$ 3.7e-5 } & \multirow{2}{7em}{\centering 3.8e-3 $\pm$ 7.7e-5 }
      & \multirow{2}{7em}{\centering 2.6e-3 $\pm$ 6.4e-5 } \\
      & & 1.1e-3 $\pm$ 2.3e-5 &  &  \\
      &  Flux ratio & 0.54 $\pm$ 0.01 & - & - & - & - \\
    \hline
    \multirow{5}{1em}{\centering Ar} &
    \multirow{2}{3em}{\centering Vz$\ $(km/s)} 
	  & -3830 $\pm$ 140 &\multirow{2}{6em}{\centering -1800 $\pm$ 80} & \multirow{2}{6em}{\centering -1460 $\pm$ 130 } & \multirow{2}{6em}{\centering -1420 $\pm$ 190 }
      & \multirow{2}{6em}{\centering -1070 $\pm$ 230 } \\
      & & 2710 $\pm$ 310 &  &  \\
    & \multirow{2}{8em}{\centering Flux$\ $(ph cm$^{-2}$ s$^{-1}$)} 
	  & 9.8e-5 $\pm$ 4.2e-6 &\multirow{2}{7em}{\centering 3.8e-4 $\pm$ 7.8e-6} & \multirow{2}{7em}{\centering 2.5e-4 $\pm$ 7.1e-6 } & \multirow{2}{7em}{\centering 3.3e-4 $\pm$ 1.4e-5 }
      & \multirow{2}{7em}{\centering 1.9e-4 $\pm$ 1.2e-5 } \\
      & & 5.9e-5 $\pm$ 5.5e-6 &  &  \\
      &  Flux ratio & 0.62 $\pm$ 0.02 & - & - & - & - \\
    \hline
    \multirow{5}{1em}{\centering Ca} &
    \multirow{2}{3em}{\centering Vz$\ $(km/s)} 
	  & -4210 $\pm$ 250 &\multirow{2}{6em}{\centering -2040 $\pm$ 120} & \multirow{2}{6em}{\centering -660 $\pm$ 180 } & \multirow{2}{6em}{\centering -900 $\pm$ 230 }
      & \multirow{2}{6em}{\centering -80 $\pm$ 470 } \\
      & & 2600 $\pm$ 330 &  &  \\
    & \multirow{2}{8em}{\centering Flux$\ $(ph cm$^{-2}$ s$^{-1}$)} 
	  & 3.0e-5 $\pm$ 2.0e-6  &\multirow{2}{7em}{\centering 1.2e-4 $\pm$ 3.2e-6} & \multirow{2}{7em}{\centering 9.1e-5 $\pm$ 3.3e-6 } & \multirow{2}{7em}{\centering 1.0e-4  $\pm$ 6.3e-6 }
      & \multirow{2}{7em}{\centering 5.7e-5 $\pm$ 5.3e-6 } \\
      & & 3.0e-5  $\pm$ 2.1e-6 &  &  \\
      &  Flux ratio & 0.50 $\pm$ 0.03 & - & - & - & - \\
    \hline
    \multirow{5}{1em}{\centering Fe} &
    \multirow{2}{3em}{\centering Vz$\ $(km/s)}
	  & -2510 $\pm$ 110 &\multirow{2}{6em}{\centering -500 $\pm$ 60} & \multirow{2}{6em}{\centering 150 $\pm$ 80 } & \multirow{2}{6em}{\centering 130 $\pm$ 140 }
      & \multirow{2}{6em}{\centering 210 $\pm$ 170 } \\
      & & 3330 $\pm$ 120 &  &  \\
    & \multirow{2}{8em}{\centering Flux$\ $(ph cm$^{-2}$ s$^{-1}$)} 
	  & 4.8e-5 $\pm$ 1.6e-6 &\multirow{2}{7em}{\centering 1.3e-4 $\pm$ 2.4e-6} & \multirow{2}{7em}{\centering 1.2e-4 $\pm$ 2.5e-6 } & \multirow{2}{7em}{\centering 1.3e-4 $\pm$ 4.8e-6 }
      & \multirow{2}{7em}{\centering 8.0e-5 $\pm$ 3.8e-6 } \\
      & & 4.2e-5 $\pm$ 1.6e-6 &  &  \\
      &  Flux ratio & 0.53 $\pm$ 0.02 & - & - & - & - \\
    \hline
    \multirow{2}{1em}{\centering Broadening} & $\sigma$$\ $(eV) & 16.5 $\pm$ 0.2 & 22.4 $\pm$ 0.1 & 23.7 $\pm$ 0.2 & 23.2 $\pm$ 0.3 & 22.3 $\pm$ 0.4 \\
     & $\Delta V_z$$\ $(km/s) & 2475 $\pm$ 30 & 3360 $\pm$ 15 & 3555 $\pm$ 60 & 3480 $\pm$ 45 & 3345 $\pm$ 60 \\
        
    \hline
    \multirow{2}{1em}{\centering Pwl} &
    index $\Gamma$ & 2.78 $\pm$ 0.04 & 2.82 $\pm$ 0.03 & 2.75 $\pm$ 0.04 &  2.83 $\pm$ 0.07 & 2.77 $\pm$ 0.06 \\
    & Flux (ph cm$^{-2}$ s$^{-1}$) & 9.2e-2 $\pm$ 5.6e-3 & 1.0e-1 $\pm$ 4.4e-3 & 9.0e-2 $\pm$ 5.2e-3 &  8.7e-2 $\pm$ 8.6e-3 & 1.3e-1 $\pm$ 1.1e-2 \\
    \hline  
\end{tabular}}\label{tab:FitGlobal}
\tablefoot{
By convention, a negative velocity corresponds to the blueshifted emission from the front side. The velocity is given in km/s. 
The reported flux corresponds to the He$\alpha$ emission line for Si and S, and the K$\alpha$ emission line for Ar, Ca, and Fe. The flux ratio parameter corresponds to the flux of the front face divided by the summed flux of the rear and front faces (measured only for the Center). The uncertainty associated with this flux ratio takes into account the covariance between the normalizations of the front and rear components, which is nevertheless small.
The flux associated with the powerlaw (pwl) is integrated over the 1.75–7 keV energy band.
The global line broadening has a linear dependency on the energy, the values (in eV) given here are defined at 2 keV. The velocity dispersion corresponding to the broadening is also indicated (in km/s). 
}
\end{table*}

\section{Results}

\subsection{Global fit}
\label{subsection:Global fit}

The first objective is to obtain the global kinematic properties of the SNR for every element, taking advantage of the full statistics per pointing. For each element (Si, S, Ar, Ca and Fe), we measure the front- and back-side velocities for the central pointing, and the line-of-sight mean velocity for each of the edge pointings (NorthWest, NorthEast, North, South). We measure the Gaussian normalizations (and deduce the ratio between front and back sides in the Center) and a global line broadening.  We also provide the powerlaw index of the continuum, whose values are consistent with previous measurements \citep[e.g., ][]{Godinaud2024}. In the remainder of this paper, we focus exclusively on the thermal emission. Table \ref{tab:FitGlobal} provides the resulting best-fit parameters. Figure \ref{fig:FirstViewSpectra} shows the fits to the global spectrum per pointing. The associated posterior distributions are well constrained and show no correlation between parameters, except for the normalization and index of the powerlaw (see Appendix \ref{Appendix : CornerPlot} for the Center pointing). Some features are observed in the residuals for some lines whose physical profiles are more complex than the Gaussian approximation. The use of a single global broadening for all lines also has limitations.
The S He${\alpha}$ line complex shows a distinct residual profile in all pointings except the center pointing, with positive residuals at the line edges and a negative residual at the line center.
This must be because the Doppler broadening due to expansion is flatter than a Gaussian.
We also observe some structures in the residuals associated with lines that we do not model, such as Mn and Cr. These lines are the subject of a dedicated study (Giuffrida et al. 2026).

Thanks to \emph{Resolve} spectral resolution, we measure the front- and back-side velocities of the shocked ejecta at the center of Tycho's SNR, as it has been demonstrated for the Fe-K (XRISM collaboration, 2026), and this for each individual intermediate mass element. 
Si and S have relatively similar line-of-sight velocities: about 3300 km s$^{-1}$ (3320 km s$^{-1}$ for Si and 3250 km s$^{-1}$ for S) for the front shell, and about 3900 km s$^{-1}$ (3920 km s$^{-1}$ for Si and 3830 km s$^{-1}$ for S) for the rear shell, revealing a velocity asymmetry.
The fits of the Si Ly$\alpha$ line (see appendix \ref{Appendix : SiLyalpha_fit}) confirm that for Si the back side is slightly faster by around 600 to 700 km s$^{-1}$, and are not biased by the choice of the rest energy. 
While for iron, the line-of-sight velocities are 2510 km s$^{-1}$ and 3330 km s$^{-1}$ for the front and rear shells respectively, indicating also a velocity asymmetry (XRISM collaboration, 2026). Si and S-rich ejecta are thus observed to be faster than those rich in Fe.
A larger velocity asymmetry is found for Ar and Ca elements. However, these lines are difficult to constrain, as they are contaminated by S lines for Ar, and Ar lines for Ca (see Figure \ref{fig:FirstViewSpectra}), in addition to being fainter. A physical plasma modeling appears thus necessary to obtain reliable measurements on Ar and Ca. 

Since both normalizations of the front and back shells are left free in our model for the central pointing, we can investigate the flux difference between the front and back shells. To quantify it, we use the ratio of the front-side normalization to the sum of the front and back sides, hence the percentage of the total line-of-sight flux originating from the front side (see Table \ref{tab:FitGlobal}). We find that the global Si flux in the Center pointing is slightly asymmetric with 57 $\pm$1\% coming from the front side which is reported here for the first time thanks to \emph{XRISM/Resolve}. This percentage decreases for S (54 $\pm$1\%) to Fe (53 $\pm$1\%).

For the global line broadening, we observe that its value is smaller at the center ($\sim$17 eV) than at the rim ($\sim$23 eV). This is consistent with the expected line-of-sight velocity dispersion: lower through the shocked shell at the center than transversely along the shocked shell at the edges. The global line broadening is slightly larger in the NorthEast ($\sim$23.7 eV) than in the NorthWest ($\sim$22.4 eV). These measurements are delicate to interpret, because we apply the same \emph{gsmooth} component for all Gaussian lines of all elements (with a linear variation of the width with energy), and for both the front and back shells at the center.

\subsection{Pixel level mapping}
\label{subsection:MappingVz}

We now study at the pixel level, and per element, the small scale properties of the dynamics and line emission.

\begin{figure*}[t!]
\centering
\includegraphics[scale=0.55, trim = 0 0 0 0, clip=true]{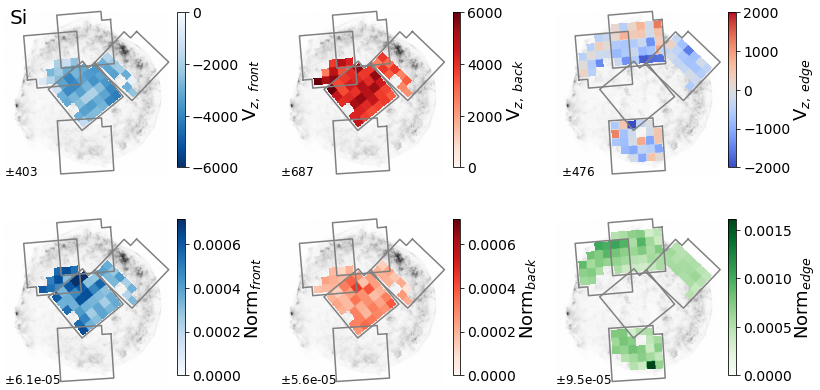}
\caption{\footnotesize Maps of line-of-sight Si velocities \emph{(top row)} and normalizations \emph{(bottom row)} for the front \emph{(first column)} and back \emph{(second column)} sides of Tycho's SNR, using the two Gaussian model, and near the edge using the one Gaussian model \emph{(third column)}. The five pointings of \emph{XRISM/Resolve} are indicated, and the maps overlaid on the 0.5-7keV \emph{Chandra} map in gray.
The average uncertainties for each of these parameters are indicated in the lower-left corner.}
\label{fig:MapSi}
\end{figure*}

\begin{figure*}[t]
\centering
\includegraphics[scale=0.29, trim = 0 0 0 0, clip=true]{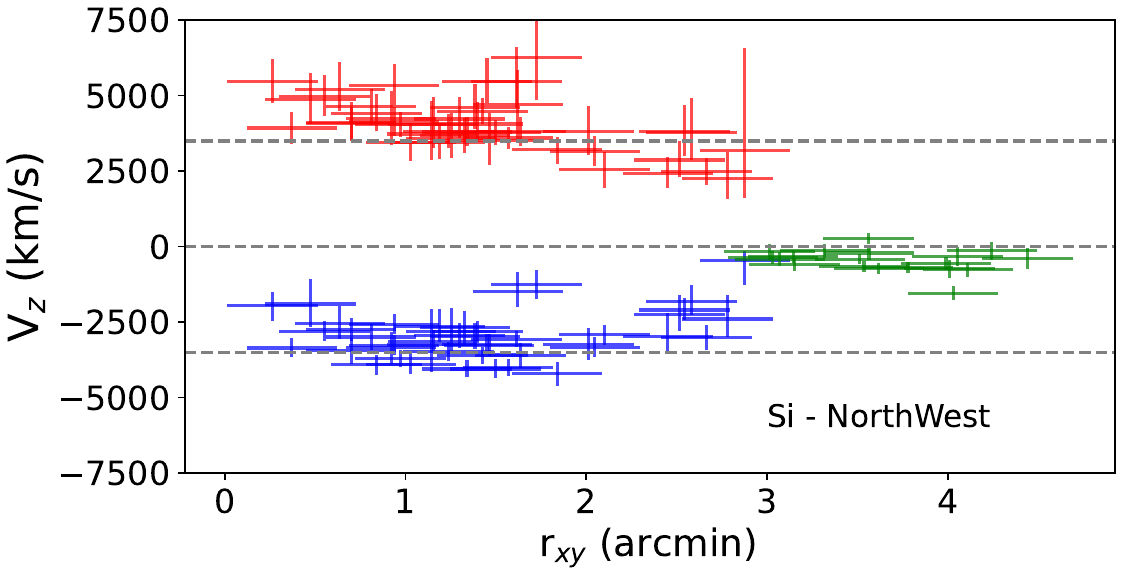}
\includegraphics[scale=0.29, trim = 0 0 0 0, clip=true]{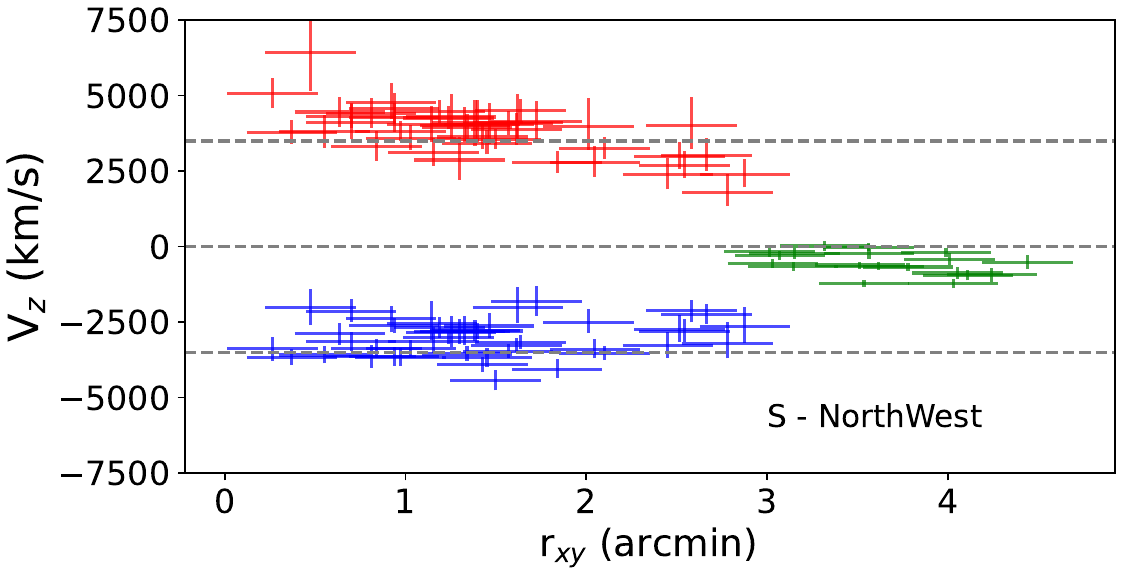}
\includegraphics[scale=0.29, trim = 0 0 0 0, clip=true]{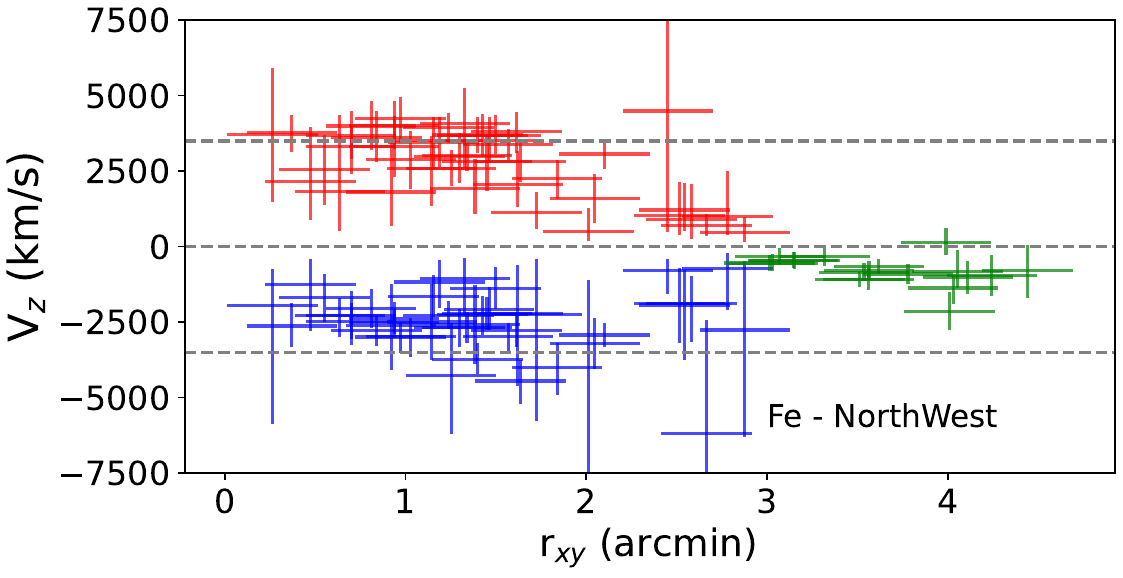}
\includegraphics[scale=0.29, trim = 0 0 0 0, clip=true]{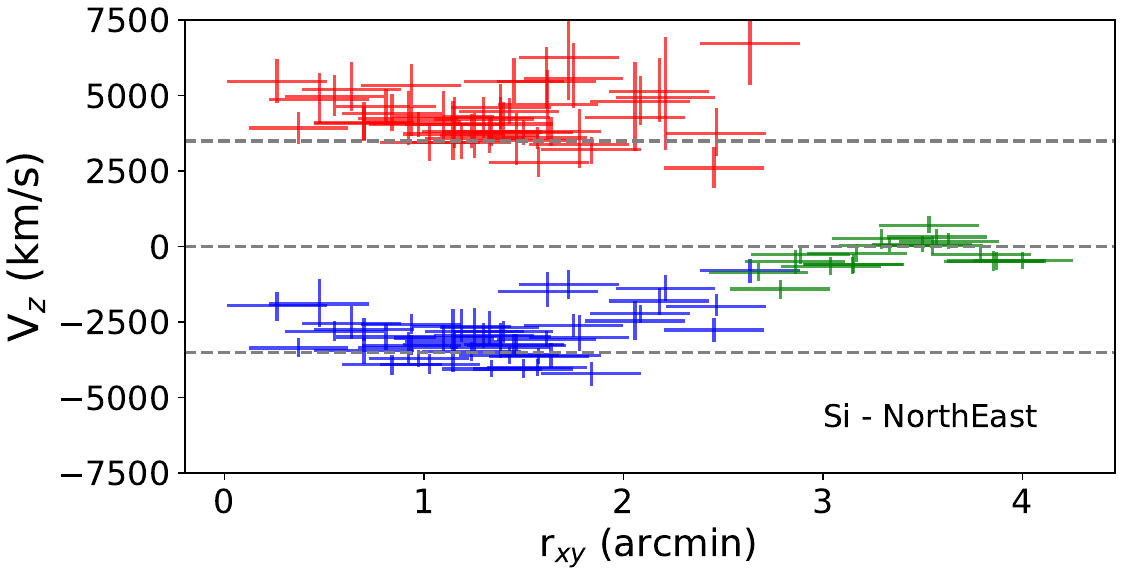}
\includegraphics[scale=0.29, trim = 0 0 0 0, clip=true]{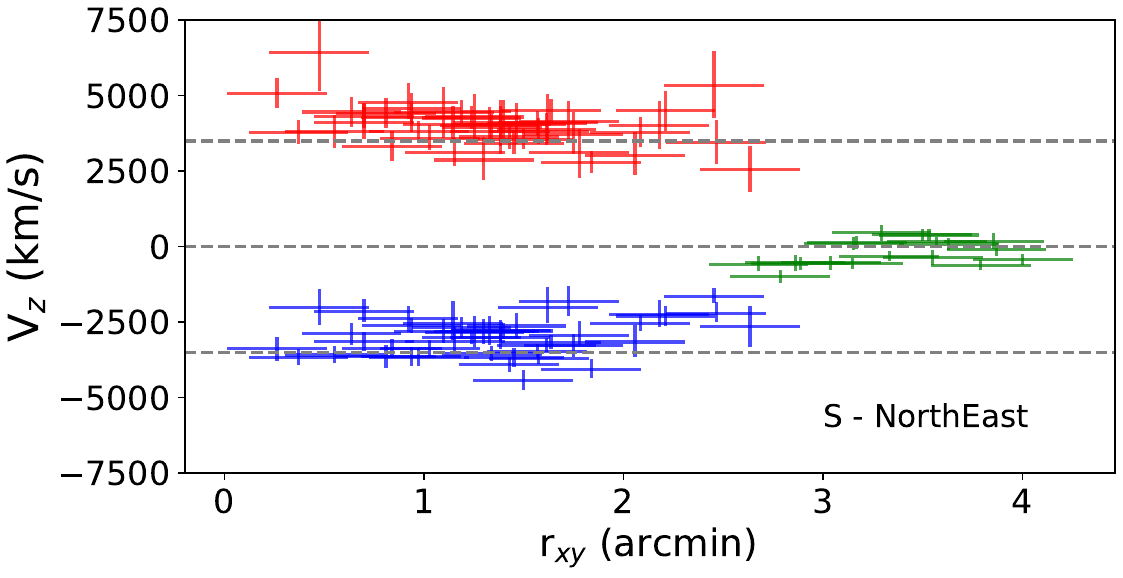}
\includegraphics[scale=0.29, trim = 0 0 0 0, clip=true]{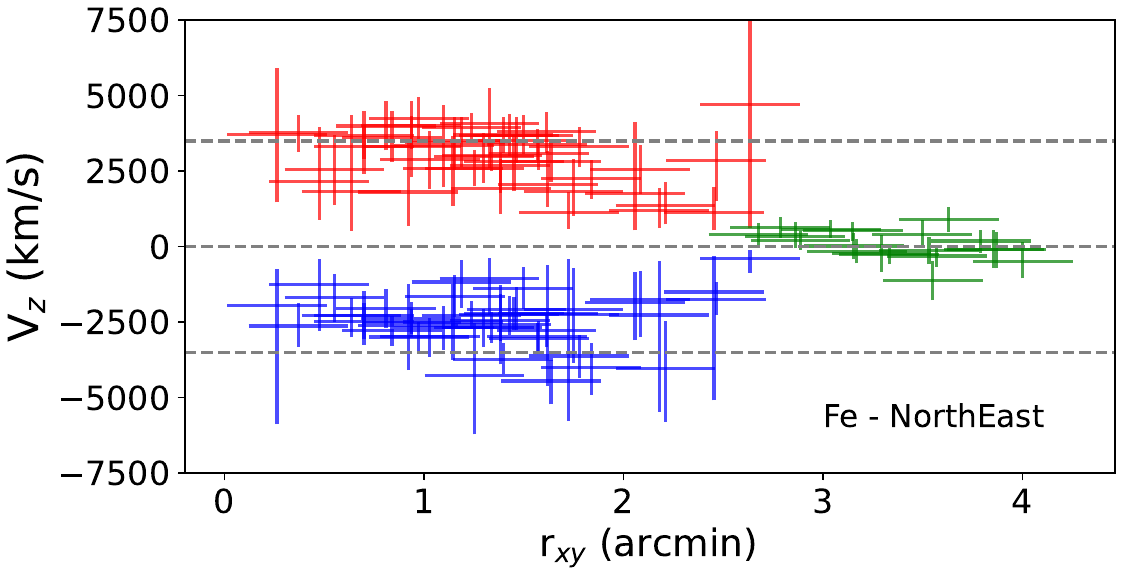}
\caption{\footnotesize Radial velocity profiles using the pixels from the Center and the NorthWest pointings (\emph{first row}) and from the Center and NorthEast pointings (\emph{second row}), for Si \emph{(left panel)}, S \emph{(middle panel)}, and Fe \emph{(right panel)} lines. Blue and red points correspond to the front and back components of the two-Gaussian model, while green points correspond to the single-Gaussian model. 
The horizontal gray dashed lines correspond respectively to -3500 km/s, 0 km/s and 3500 km/s.}
\vspace{-0.2cm}
\label{fig:Profileradiaux}
\end{figure*}

\begin{figure}[t]
\centering
\includegraphics[scale=0.37, trim = 0 0 0 0, clip=true]{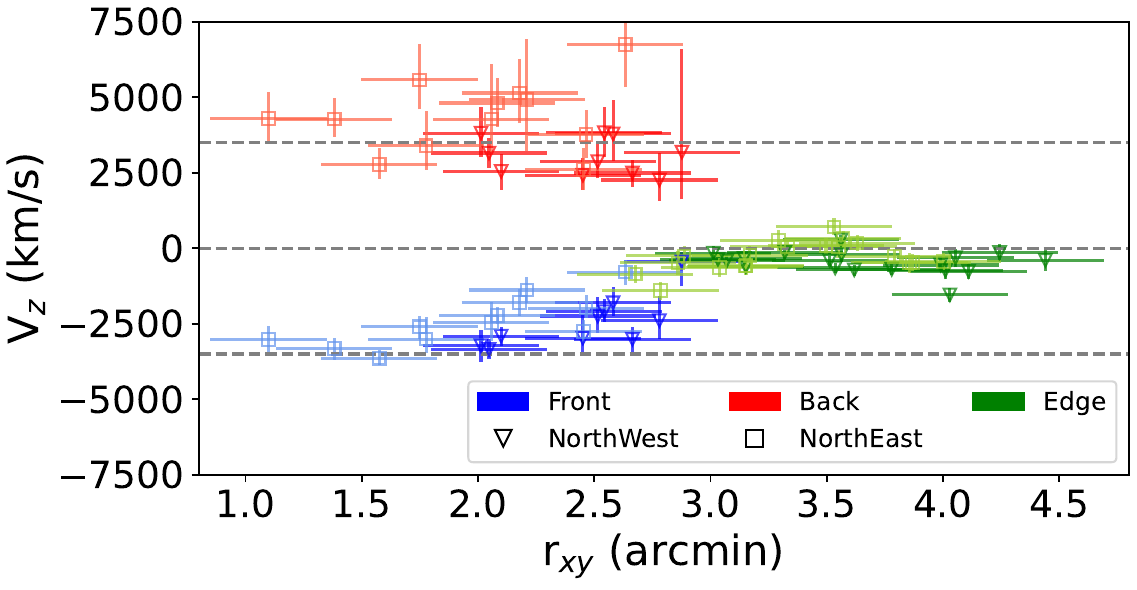}
\caption{\footnotesize Radial profile of the Doppler velocity $V_{z}$ as a function of the radius $r_{xy}$ for the pixels of the NorthWest (triangle marker and darker color) and NorthEast (square marker and lighter color) pointings for Si. A slight azimuthal variation between these two pointings can be observed between 2 and 2.5 arcmin.}
\label{fig:Profileradiaux_NW-NE}
\vspace{-0.2cm}
\end{figure}

\begin{figure*}[t!]
\centering
\includegraphics[scale=0.27, trim = 80 310 80 340, clip=true]{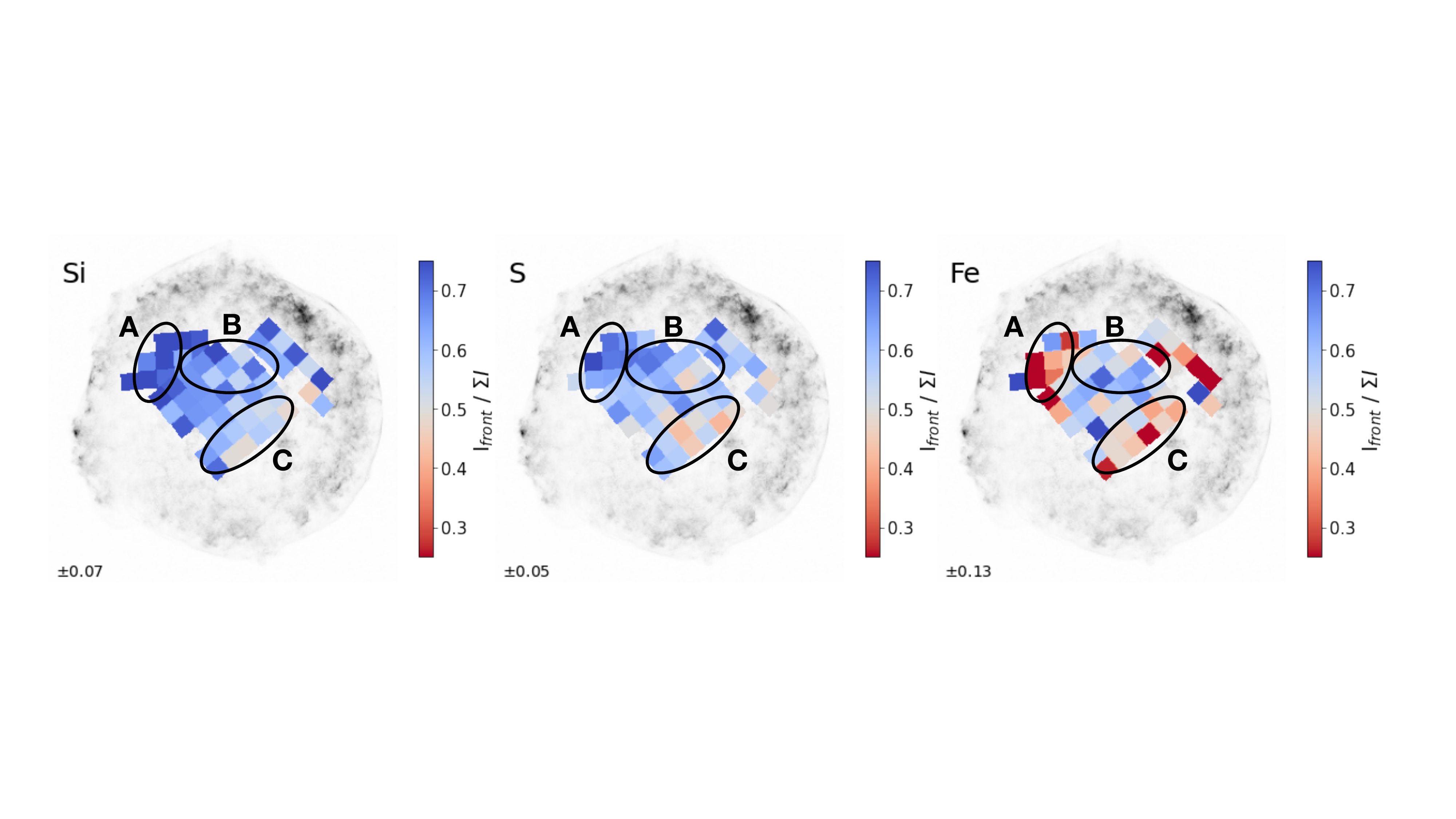}
\caption{\footnotesize Ratio between the normalization of the front face and that of the sum of the rear and front faces. This parameter can only be calculated where we use the two-Gaussian model to measure the properties of the front and rear faces separately. This map is presented for Si \emph{(left)}, S \emph{(middle)} and Fe \emph{(right)}. On these maps are overlaid three zones to describe the results in the text. The average uncertainties for each of these parameters are indicated in the lower-left corner.}
\vspace{-0.2cm}
\label{fig:CarteRatioNorm}
\end{figure*}

\begin{figure*}[t]
\centering
\includegraphics[scale=0.45, trim = 0 0 0 0, clip=true]{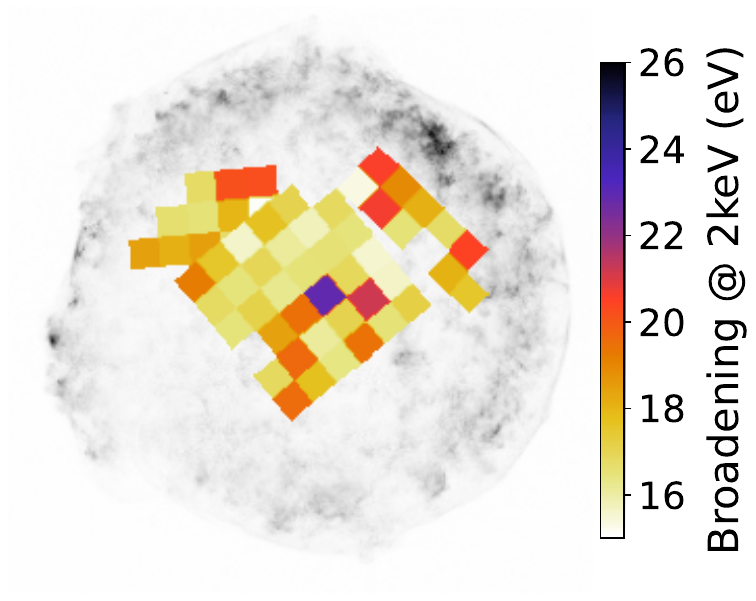}
\includegraphics[scale=0.45, trim = 0 0 0 0, clip=true]{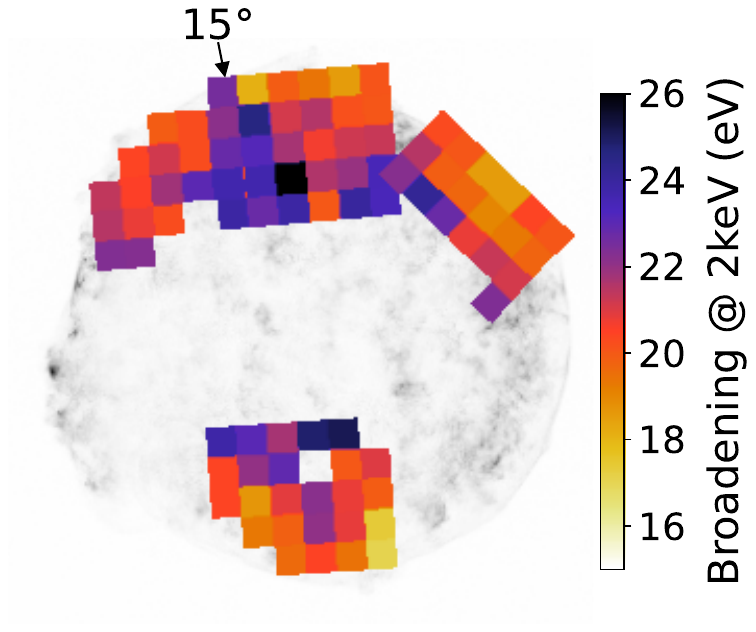}
\includegraphics[scale=0.4, trim = 21 0 0 0, clip=true]{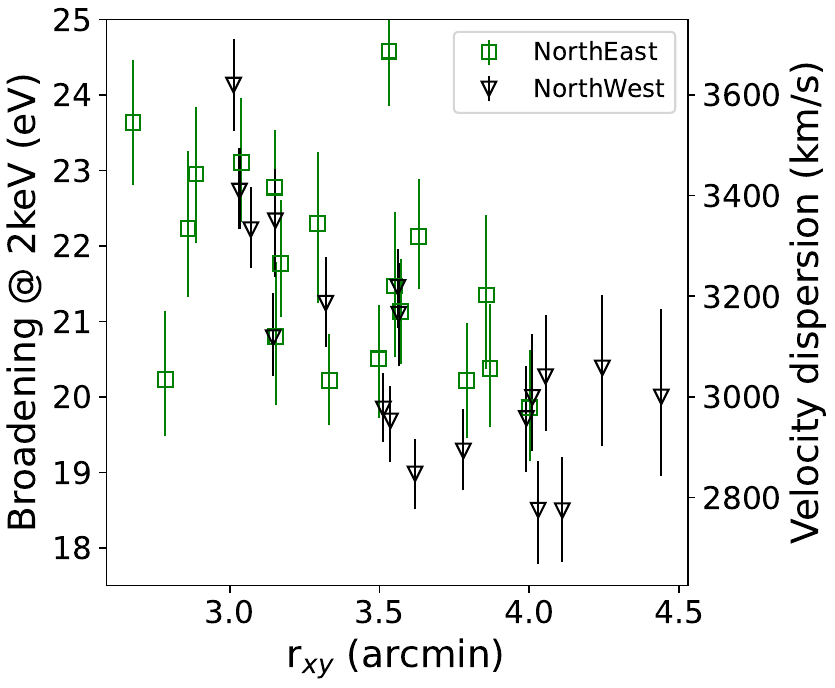}
\caption{\footnotesize \emph{Left:} Global line broadening map obtained with the \emph{gsmooth} component applied to the all emission lines in each pixel from the front and back sides. The map only includes the pixels fitted using the two-Gaussian model. \emph{Center:} The global line broadening map for the pixels fitted using the single-Gaussian model. An interesting structure is highlighted at 15°. 
\emph{Right:} Radial profile of the global line broadening for the NorthWest (black triangles) and NorthEast (green squares) pixels, using the single Gaussian model.
Since this broadening has a linear dependency on energy, the values reported in these three figures correspond to a line at 2 keV.}
\label{fig:MapBroadening}
\vspace{-0.2cm}
\end{figure*}

\subsubsection{Front and back line-of-sight velocities}
\label{subsection:Results-Vz}

Figure \ref{fig:MapSi} presents the Si line velocity maps (top panels). They show where two Doppler components are required and where a single Doppler component is sufficient to reproduce the emission lines. The left and middle panels show the velocity towards the center of the remnant from the front (blueshifted) and rear (redshifted) shells, respectively. The right panel provides the mean velocity map along the shell seen transversely at the edge of the remnant. Their respective line normalization maps are shown in the bottom panels. 
These maps show some variations in velocity and flux on the scale of \emph{Resolve} pixels.
To further investigate the dynamical properties of the ejecta, Figure \ref{fig:Profileradiaux} shows the radial profiles of the Doppler velocity for Si, S and Fe from the center of the remnant to the northwestern and northeastern edges (top and bottom panels, respectively). 
Appendix \ref{Appendix : Radial profile} completes these profiles for the Ar and Ca velocities. The North and South pointings have too low exposure time and insufficient statistics to separate the front and back sides. 
The first important observation is that the front and rear sides of the remnant are clearly separated for every element and all pixels at the center. Such a clear line splitting has not yet been observed in any other young remnant with \emph{XRISM/Resolve}. 

The radial velocity profiles (Figure \ref{fig:Profileradiaux}) further confirm the finding from the global spectrum analysis that the line-of-sight dynamics of Tycho’s SNR is asymmetric. 
Figure \ref{fig:Profileradiaux_NW-NE} presents the comparison of the radial Si velocity profile between the northwestern and northeastern pointings. While the general behavior is similar, there is some indication that the blueshifted side is slower between 2 and 2.5 arcmin (and the redshifted side faster, although with larger error bars) in the NorthEast than in the NorthWest pointing. However, the error bars on the radius are large due to the size of the \emph{Resolve} pixels.
It is interesting to note that, in the NorthEast, there is evidence in the literature that the SNR is currently interacting with a molecular cloud \citep{Zhou2016, Arias2019, Bo2026}.

The line-of-sight velocity, shown in green in radial profiles (Figures \ref{fig:Profileradiaux} and \ref{fig:Profileradiaux_NW-NE}), corresponds to cases where a single-Doppler component reproduces the lines. In some pixels of the North and South pointings, it can be due to a lack of statistics. Near the rim,  the emission is expected to result from a continuous range of shocked ejecta, as the line of sight spans transversely the shocked shell from blueshifted to redshifted material. The derived velocity is thus a mean line-of-sight flux-weighted velocity from the front to rear side of the remnant. This velocity appears slightly blueshifted at the edge (NW and NE) for Si, which was suggested from previous CCD measurements \citep{Godinaud2023, Uchida2024, Godinaud2024}. It may be due to a bulk motion of the remnant as a whole, but also may result from a flux asymmetry at the edge, or an ionization effect shifting the rest energy with regards to the assumed value for our Gaussian model.

\subsubsection{Flux asymmetry between the front and rear sides}
\label{subsection:Results-Flux}

The Gaussian normalization maps are presented for the Si emission in Figure \ref{fig:MapSi} (bottom panels). The Doppler line separation allows for the first time to disentangle the emission from the front and back sides towards the center of the remnant. The maps reveal a fainter Si emission from the redshifted than from the blueshifted sides of the remnant, and confirm the global flux Si asymmetry found for the Center global spectrum (see sect \ref{subsection:Global fit}, Table \ref{tab:FitGlobal}). 

Figure \ref{fig:CarteRatioNorm} shows the flux-ratio maps $R=\frac{I_{front}}{I_{front} + I_{back}}$ for Si, S, and Fe. 
These flux-ratio maps reveal a spatial variation (in the plane of the sky) of this front/back flux asymmetry. 
Looking by element, we observe that for Si, the flux mainly comes from the front side, particularly in the NE/N (zones A and B), and equalizes with the back side in the SouthWest (zone C), going typically from a flux ratio of 0.8 to above 0.5. The same trend is observed for S, going from about 0.7 (NE) to below 0.5 (SW). For Fe, the spatial variation of the flux ratio is clearly different: the rear side is brighter in the NE and SW (values around 0.3), and the front side slightly brighter in zone B (0.6).

\subsubsection{Global line broadening}
\label{subsection:Results-broadening}

\begin{figure*}[t!]
\centering
\includegraphics[scale=0.37, trim = 0 0 0 0, clip=true]{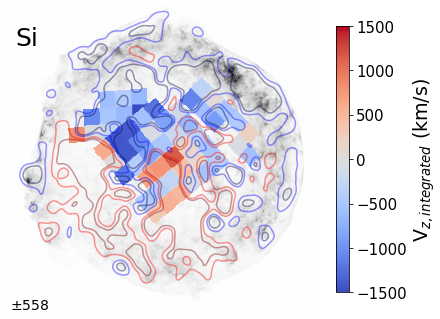}
\includegraphics[scale=0.37, trim = 0 0 0 0, clip=true]{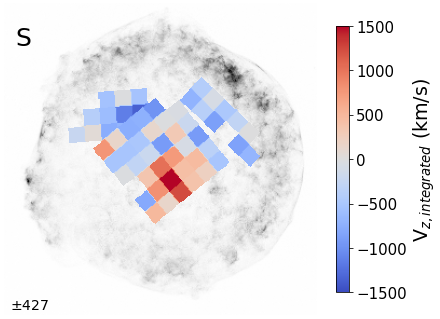}
\includegraphics[scale=0.37, trim = 0 0 0 0, clip=true]{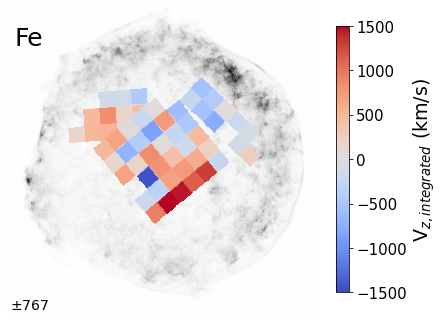}
\caption{\footnotesize
The line-of-sight mean velocity reconstructed with XRISM, equivalent to what is measured at CCD resolution.
This parameter is calculated and shown at the center, where we can measure the properties of the front and rear sides separately. This map is presented for the Si \emph{(left)}, S \emph{(middle)} and Fe \emph{(right)}. On the Si map are overlaid the contours from the integrated line-of-sight mean velocity obtained with \emph{Chandra} data for the Si line \citep{Godinaud2023}. The average uncertainties for each of these parameters are indicated in the lower-left corner.}
\label{fig:CarteVint_XRISM_Chandra}
\vspace{-0.2cm}
\end{figure*}

We evaluate the global line broadening by applying a single \emph{gsmooth} to all emission lines, with a linear energy dependence. Figure \ref{fig:MapBroadening} shows the resulting global line broadening map for the pixels fitted using the two-Gaussian model (left panel), the map for the pixels fitted using the single-Gaussian model (center panel) and the radial profile for the NorthWest and NorthEast rims (right panel). With the pixel-level mapping, we confirm the trends observed in the fits of the global spectra (see Table \ref{tab:FitGlobal}). The global line broadening at the center is around 18 eV for a line located at 2 keV, corresponding to a velocity dispersion of 2700 km/s in each half shell. At the edges, the single-Gaussian model is used, and the broadening decreases from 24 eV (corresponding to a velocity dispersion of 3600 km/s) to about 20 eV (equivalent to 3000 km/s) for the outermost pixels.

We interpret our map of global line broadening as being mostly driven by the velocity dispersion along the line of sight. Its radial evolution matches qualitatively the expected velocity dispersion for a spherical shell in expansion. At the center, the broadening corresponds to the velocity dispersion in each half shell from the front and back sides. Then these two half-shells join at the position of the reverse shock. Near the edges, for a radius larger than the reverse shock, the line of sight crosses continuously and transversely the shocked shell. The broadening decreases as expected toward the outer edge of the spherical shell, where the ejecta layer becomes thinner.

As already suggested by the global-pointing analysis, the line broadening map shows slightly larger values in the NorthEast than in the NorthWest. We identify a larger broadening structure around 15°, as marked in Fig. \ref{fig:MapBroadening} (left panel), corresponding to peculiar dynamical measurements. At 15°, \cite{Godinaud2023} identified a local slowdown in the radial proper-motion profile. The northeastern rim is indeed known to be interacting with a density enhancement in the ambient medium \citep{Zhou2016, Arias2019, Bo2026}. 

However, the use of a single \emph{gsmooth} component for all the Gaussian lines comes with several limitations for deriving the broadening :  same broadening for all elements and for all lines (e.g., He$\alpha$ and He$\beta$, see Figure \ref{fig:AtomDB}), no physical modeling of the intrinsic line complex width. Finally, the measured broadening is affected by spatial–spectral mixing (SSM) due to the XRISM PSF, and each pixel is contaminated by neighboring pixels. We do not account for the effects of the SSM in this study. More advanced methods, such as data fusion \citep{Lascar2025} combining high spectral resolution observations (e.g., \emph{XRISM/Resolve}) with higher spatial resolution data from \emph{Chandra} and \emph{XMM-Newton}, could allow for a better treatment of the SSM in the future.

\section{Discussion}

The \emph{XRISM/Resolve} observations of Tycho’s SNR have revealed the line-of-sight kinematics of the ejecta, and that per element, by mapping at pixel level the variations of the Doppler shifts from the center to the rim. Towards the center of the remnant, they provide an independent measurement of the velocity in the front and rear shocked shells.
For Si and S, the velocity range is asymmetric from about -3300 km/s towards us and 3900 km/s on the rear side. Fe shows lower velocity values but also asymmetrical, around -2500 km/s for the front shell and 3300 km/s for the rear shell, and slightly less asymmetry.
In addition, the Si flux exhibits a significant asymmetry between the front and rear shells, varying spatially from more than 70 $\%$ front-dominated towards the northeast rim down to 50 $\%$ towards the southwestern rim. This asymmetry is observed as well for S, with a stronger variation from the NE (70 $\%$ front-dominated) to the SW, where it is rather rear-dominated (up to 60 $\%$). The Fe front-to-total flux ratio map appears more patchy, with 70 $\%$ predominant emission from the rear shell towards both the NE and SW rims, with a rather front-dominated center region.

\subsection{Comparison with previous CCD studies}
\label{subsection:Disct_FluxAssym}

\begin{figure*}[t]
\centering
\includegraphics[scale=0.36, trim = 150 505 320 120, clip=true]{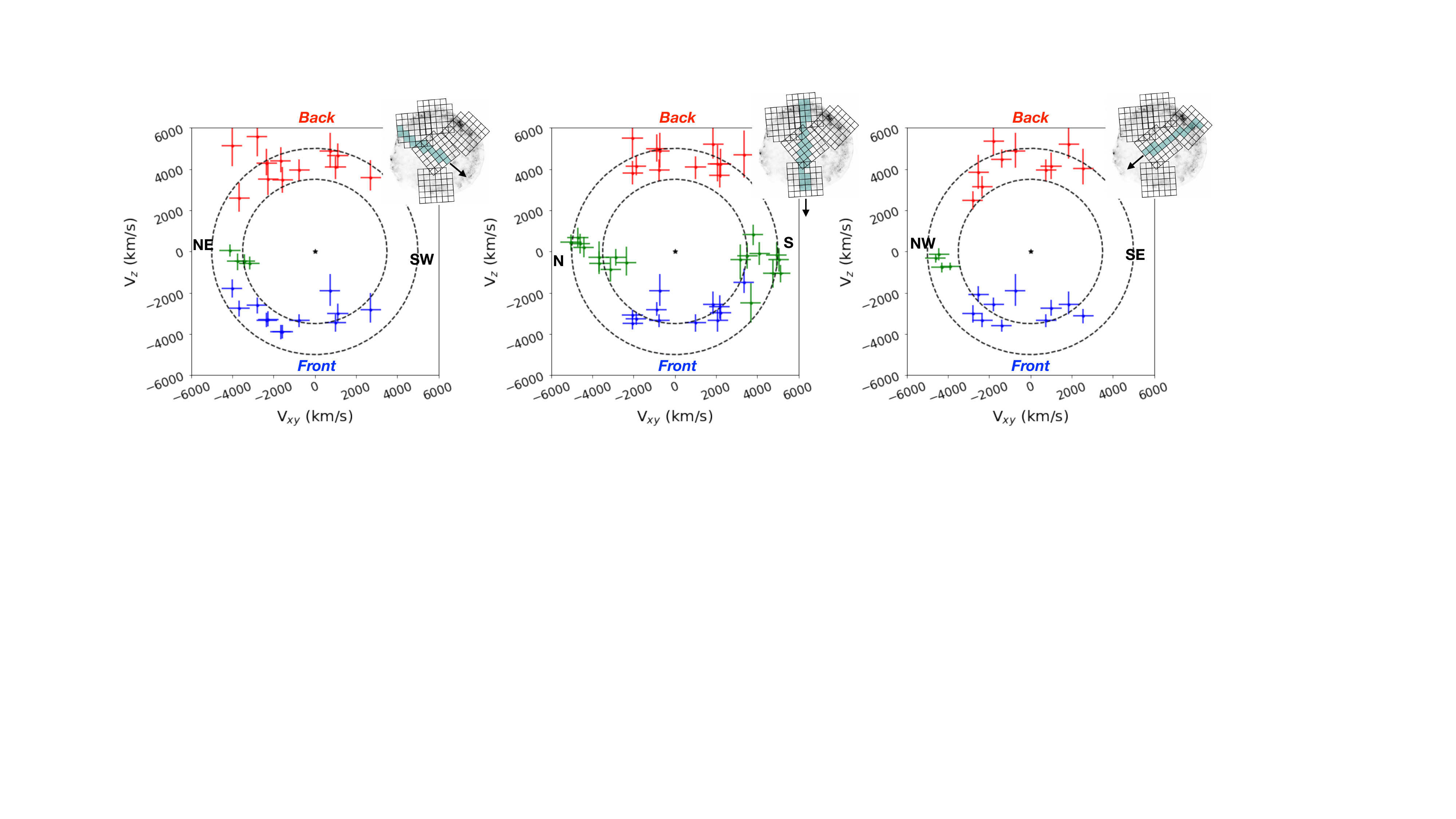}
\caption{\footnotesize 3D reconstruction in velocity space of the Si dynamics, for the three slices along the axis indicated in blue in the associated image (upper right). The velocity in the plane of the sky, V$_{xy}$, is derived from proper motion measurements using \emph{Chandra} data \citep{Godinaud2023}. The Doppler velocities, V$_{z}$, are measured in this analysis using \emph{XRISM/Resolve} data. The red (back side) and blue (front side) points correspond to fits with two Gaussian components, while the green points represent single-Gaussian fits near the edges of the SNR.
The black dashed circles correspond to a spherical 3D expansion with velocities of 3500 and 5000 km/s and are centered on 0 (the star marker).}
\label{fig:Reconstruction3D}
\vspace{-0.2cm}
\end{figure*}

To determine the line-of-sight mean velocities across the remnant, previous studies with CCD detectors measured the Doppler shift either by fitting the spectrum of individual bright and dense knots \citep{Sato2017a, Williams2017, Millard2022}, and larger regions over the whole remnant \citep{Hayato2010, Uchida2024, Godinaud2024, Petruk2025} or by reconstructing the energy map of the Si line \citep[][see Figure \ref{fig:PointingXRISM}]{Godinaud2023}. As a result, the northern part of Tycho’s SNR appeared globally blueshifted, while the southern part is mostly redshifted.
With the \emph{XRISM/Resolve} observations, we independently measure for the first time the properties of the front and back sides, and observe both significant velocity (Section \ref{subsection:Results-Vz}) and flux (Section \ref{subsection:Results-Flux}) asymmetries.
This leads us to re-evaluate the interpretation of the CCD-resolution results. Indeed, the choice of bright knots could be biased by the flux asymmetry, and the measured Doppler velocity is in fact an integrated velocity weighted by flux along the line of sight, what we call in this study the "line-of-sight mean velocity".

We thus produce a map of the line-of-sight mean velocity $V_{z,mean}$ from our \emph{XRISM/Resolve} measurements to compare with previous CCD results. This quantity can be expressed to first order as:
\[ V_{z, mean} = V_{z,front} \times \frac{I_{front}}{ I_{front} + I_{back}} + V_{z,back} \times (1-\frac{I_{front}}{ I_{front} + I_{back}}) \]

Figure \ref{fig:CarteVint_XRISM_Chandra} shows the reconstructed map of $V_{z,mean}$ for Si, S, and Fe from our \emph{XRISM/Resolve} measurements. Overlaid on this figure are the contours of the line-of-sight mean velocity map (see Figure \ref{fig:PointingXRISM}) obtained with \emph{Chandra} for the Si line \citep{Godinaud2023}. The contours levels correspond to red- or blueshifted mean velocities of 500 and 1000 km/s, with an 8-arcsec smoothing applied to the \emph{Chandra} map. We observe a good correlation between these two independent measurements of $V_{z,mean}$ for Si and S. For Fe, some structures are consistent, but this is not systematically the case, which is  expected since the \emph{Chandra}-derived contours correspond to Si.

The normalization ratio map shown in Fig. \ref{fig:CarteRatioNorm} corresponds to the term $\frac{I_{front}}{ I_{front} + I_{back}}$ in the expression of $V_{z,mean}$. In these flux-ratio maps, we observe structures similar to those seen in the $V_{z,mean}$ maps: the northern region -which appears blueshifted in \emph{Chandra} integrated map- has a brighter front face, especially for Si. The southern region shows a comparable flux ratio between the front and back sides for Si, hence the line-of-sight mean velocity is redshifted because the back side is faster. 
The flux asymmetry combined with the velocity asymmetry that we observe for the first time with \emph{XRISM/Resolve} therefore has a major impact on the interpretation of previous CCD results.

\subsection{3D reconstruction of the Si ejecta dynamics}
\label{subsection:Disct_Vz3D}

Thanks to the Doppler velocity (V$_{z}$) measurements of the front and rear sides which cover a significant fraction of the remnant, it is now possible to reconstruct a 3D velocity map by combining them with proper motion (V$_{xy}$) measurements in the plane of the sky.
Such studies were previously done for the bright Si ejecta knots of Tycho's SNR \citep{Williams2017, Millard2022}. With \emph{XRISM/Resolve}, we can access the 3D dynamics of Tycho's SNR with independent Doppler velocity measurements for the front and rear sides, and over a large fraction of the SNR.

For this, we use the proper motions of 1700 small ejecta features, derived from the 2003 and 2009 broadband \emph{Chandra} observations (dominated by IMEs) of Tycho’s SNR \citep{Godinaud2023}. For each pixel of the \emph{XRISM/Resolve} pointings, we compute the average of the vectors located at that position, assuming the same V$_{xy}$ for both the front and rear sides of the remnant. We then create azimuthal slices in the (V$_{xy}$, V$_{z}$) plane to visualize the 3D Si dynamics, using V$_{z}$ measured for the front and rear sides of the shocked ejecta shell. 
The V$_{xy}$ velocities assume a distance of 3.5 kpc \citep{Hayato2010, Williams2017}. V$_{xy}$ scales linearly with the distance, whereas V$_{z}$ does not depend on it. The front/rear velocity asymmetry and the flux asymmetry are therefore independent of the assumed distance, only the comparison with V$_{xy}$ is affected. Using the more recent and more precise distance of 3.2$^{+0.1}_{-0.2}$ kpc found by \cite{Neumann2024} from light echoes, the V$_{xy}$ velocities would be reduced by $\sim$8.6\%.

Figure \ref{fig:Reconstruction3D} presents the results for three slices along the NE/SW, N/S and NW/SE axis. This representation allows one to compare directly the velocity properties in the plane of the sky and along the line of sight. It shows kinematic asymmetries in 3D, which might be related to overdensities in the surrounding medium. 
Along the axis from the NE to center (Fig. \ref{fig:Reconstruction3D}, left panel), we observe a velocity indentation at the NE shock, with some higher velocities from the front/rear ejecta, than radially.
This slowdown might result from the presence of a possible overdensity in the ambient medium observed in radio \citep{Zhou2016, Bo2026}. Moving along the axis towards the center, we observe a slower front side velocity, similar to that of the NE slowdown. This might indicate interaction of the front side with a locally higher density surrounding medium, although this area is not known to be interacting with the surrounding medium. Radio and infrared observations do not indicate any particular overdensity there. Along the N/S axis (Fig. \ref{fig:Reconstruction3D}, middle panel), the velocity is higher at the rear than front side of the ejecta shell. Similarly high velocities are found at the outermost north and south rims, unlike at the NE. A similar behavior is observed along the NW /center axis (Fig. \ref{fig:Reconstruction3D}, right panel). 

The velocity asymmetry between the front and rear sides is moderate ($\sim$16\% for Si) and of the same order as the azimuthal variations of the expansion velocity observed in the plane of the sky \citep{Katsuda2010, Williams2016, Godinaud2023}. It reflects the recent dynamics and does not necessarily imply a geometrical line-of-sight asymmetry of the remnant, whose radio and X-ray morphology appears rather round. A slightly denser medium on the front side could slow it down without significant distortion of the shell, if the interaction happened recently, for example in the case of a density gradient in the surrounding medium \citep{Williams2013, Williams2016}.

\subsection{Element stratification}
\label{subsection:Disct_elements}

\begin{figure}[t]
\centering
\includegraphics[scale=0.55, trim = 0 0 0 0, clip=true]{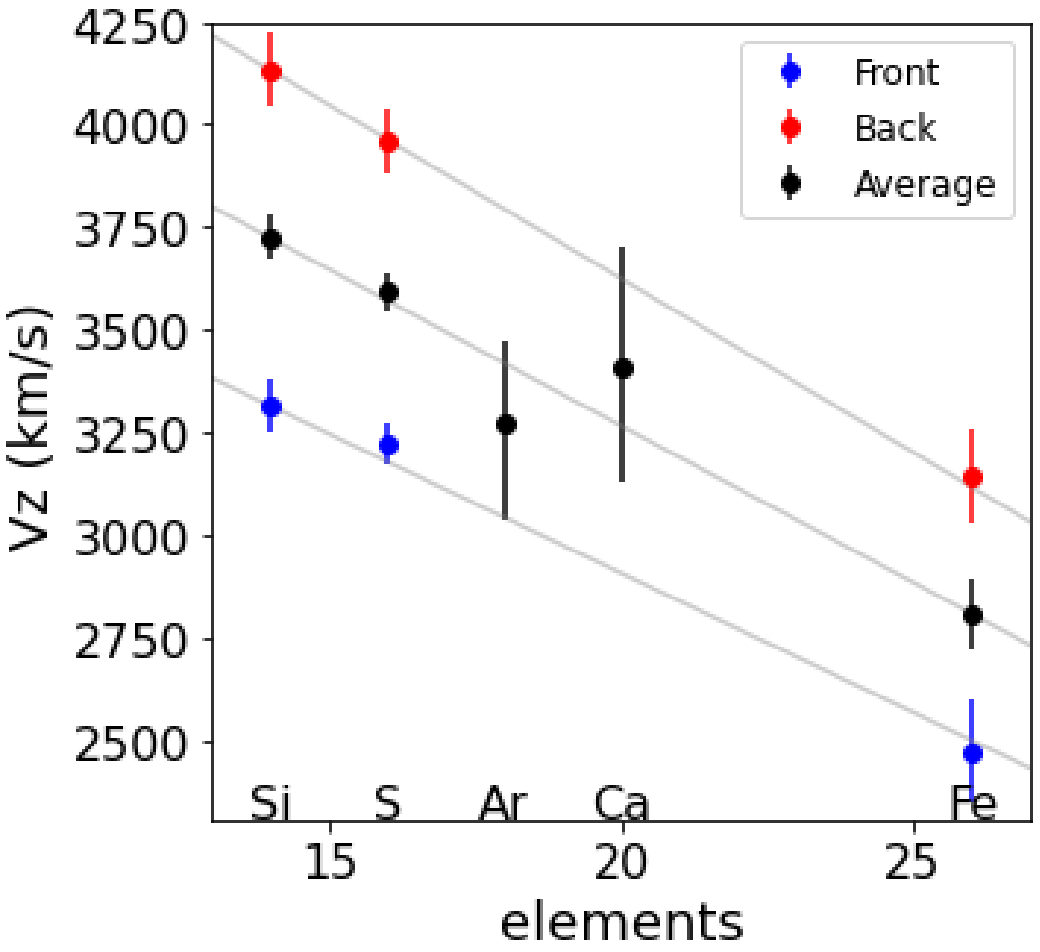}
\caption{\footnotesize
The mean absolute value of the Doppler velocities (using all pixels from the Center pointing) of the front side (blue), rear side (red), and the average of both sides (black) as a function of the proton number for the elements studied here: Si, S, Ar, Ca, and Fe. The gray lines show the velocities obtained under the assumption that the velocity ratio between each element and Si, $R(Z) = V(Z)/V(\mathrm{Si})$, is identical for the three velocity sets (front, rear, and average). The ratio $R(Z)$ is described by a linear relation, $R(Z)=aZ+b$, with $a$ and $b$ determined from the average velocities.}
\label{fig:StratificationElt}
\vspace{-0.2cm}
\end{figure}

The \emph{XRISM/Resolve} observations of Tycho's SNR provide an unprecedented 3D view of the shocked ejecta and allow us to derive the kinematics of individual elements. This enables the study of the velocity stratification of the elements and their physical properties inside the ejecta-emitting shell. 
The morphology of the remnant in specific emission lines is often used to search for any element stratification in the ejecta. In Tycho’s SNR, the Fe-K emission appears radially inside that of Si \citep{Decourchelle2001, Hwang2002}. 
The ratio between the mean radii of Fe-K ($\approx$ 170") and Si ($\approx$ 200") is about 0.85 \citep[see Figure 2 and 3,][]{Decourchelle2001}. 
The spectral properties of the non-equilibrium ionization plasma also exhibit a stratification \citep{Badenes2006, Yamaguchi2017}, with the ionization timescale decreasing with increasing radius.
While morphological studies provide information on the mean radial stratification of the elements in the plane of sky, Doppler velocities measurements inform us on the level of stratification of the elements in the line of sight. Assuming spherical expansion of the SNR, \cite{Hayato2010} used \emph{Suzaku} observations to estimate velocities per element, based on the line broadening per element and versus radius. Assuming same front/rear velocities, they obtained a Fe velocity of 4000 $\pm$ 300 km s$^{-1}$ also lower than that of the IMEs (4700 $\pm$ 100 km s$^{-1}$). 
\cite{Petruk2025} also compared the line-of-sight mean velocities of Si and S using Chandra data, by fitting their lines positions with a Gaussian across the SNR. Their result for Si is similar to that of \cite{Godinaud2023}, but their S map is much more redshifted, leading them to conclude that Si and S have different properties. This difference, however, may be due to the choice of the rest energy of the S line, which does not account for non-equilibrium ionization effects.

With \emph{XRISM/Resolve}, we separately measure the front and back velocities of the shocked ejecta and reveal that the expansion of Tycho's SNR is not spherical, and has asymmetric front/rear velocities along the line of sight. We find smaller values of the velocities than derived previously from CCD observations from \citet{Hayato2010}.
In Figure \ref{fig:StratificationElt}, we show the mean velocities (averaging all pixels from the central pointing) for each element and separately for the front (in blue) and rear (in red) sides. In black, we add the mean front/rear velocity, which is independent of the chosen reference energy (provided that the front/rear plasma are in the same physical conditions), thus removing some bias for Ar and Ca (see Appendix \ref{Appendix : Eref}).
We find a linear decrease in the velocity with increasing atomic number, from Si to Fe. 
Despite the front side velocities being lower than those of the rear side, the velocity ratios between elements are kept similar. This ratio has a value of 0.75 between  Fe and Si, comparable to the Fe/Si radius ratio of roughly 0.85 measured in the plane of the sky.
The velocity profile as a function of radius within the shocked ejecta depends on the history of the reverse shock velocity, which itself depends on the initial density profile and the evolutionary stage of Tycho. The lowest velocities in the shocked ejecta are located at the contact discontinuity for power-law density models \citep{Chevalier1982}, and at the reverse shock for exponential density profiles \citep{Dwarkadas1998}. Tycho’s observations  are more consistent with an exponential initial density profile in the ejecta, with iron located inside Si and S and having a lower velocity than Si and S.

This information provides constraints to the explosion mechanism of Tycho's supernova, in addition to abundance measurements of each element and their comparison with nucleosynthesis yields from numerical simulations of different explosion mechanisms \citep[e.g., see Figure 7 in][]{Fujimaru2026}.
Element stratification is also expected by numerical simulations of Type Ia supernova explosions and varies significantly between models. For example, pure-deflagration models predict no stratification between elements \citep{Fink2014}. They are thus rejected for Tycho's SNR. Delayed-detonation models predict an onion-like structure with Fe at the center, followed by increasingly lighter elements at larger radii \citep{Seitenzahl2013}. Double-detonation models produce a very particular structure, with an inner Ca layer that is close to Fe due to the second detonation, and an outer layer at a radius larger than Si due to the first detonation \citep{Tanikawa2018}. This double structure has been observed for the first time in the optical for a SNR in the Large Magellanic Cloud \citep{Das2025}.
Our results on the stratification provide additional support on Tycho's SNR originating from either delayed detonation or double detonation models, both currently viable to describe the properties of this SNR \citep{Badenes2006, Holland-Ashford2025}.

The average velocities of Ar and Ca are also shown in Figure \ref{fig:StratificationElt}. 
They are compatible with the same trend, but cannot confirm it due to larger uncertainties.
These elements are particularly difficult to study for various reasons. They have low emission line statistics, their energy at rest is difficult to define (see Appendix \ref{Appendix : Eref}) and their associated lines are contaminated by the neighboring elements: an Ar line is emitted close to the Ca line, and a S line contaminates the Ar line. We accounted for this in our model using Gaussians, but an approach with a more physical model could improve this measurement.
In previous studies of Tycho's SNR, the properties of Ca have often been difficult to constrain or have been found to be inconsistent with those derived from the other detected elements \citep{Badenes2006, Hayato2010, Picquenot2024}.
The Ca measurement appears also difficult to constrain in \cite{Agarwal2026}, which uses physical fits for Cas A.
The large error bars do not allow us to draw conclusions about the velocity of Ca relative to the other elements, but this measurement would provide strong clues to discriminate between the delayed-detonation and double-detonation models for Tycho's SNR. The study of Ca in this SNR could also shed new light on the High Velocity Features observed in supernova spectra shortly after the explosion, particularly for Ca and Si lines \citep{Mazzali2005}. Their origin is still unclear: they could be related to Ca located in the outer layers, in dense and fast knots, or due to interaction with a particular CSM \citep{Silverman2015, Clark2021}. One of these high-velocity features was detected in the light-echo spectrum of the supernova that produced Tycho's SNR \citep{Krause2008}.

\subsection{Three-dimensional interpretations}
\label{subsection:Disct_3DInterpretation}

\begin{figure}[t]
\centering
\includegraphics[scale=0.37, trim = 300 190 300 110, clip=true]{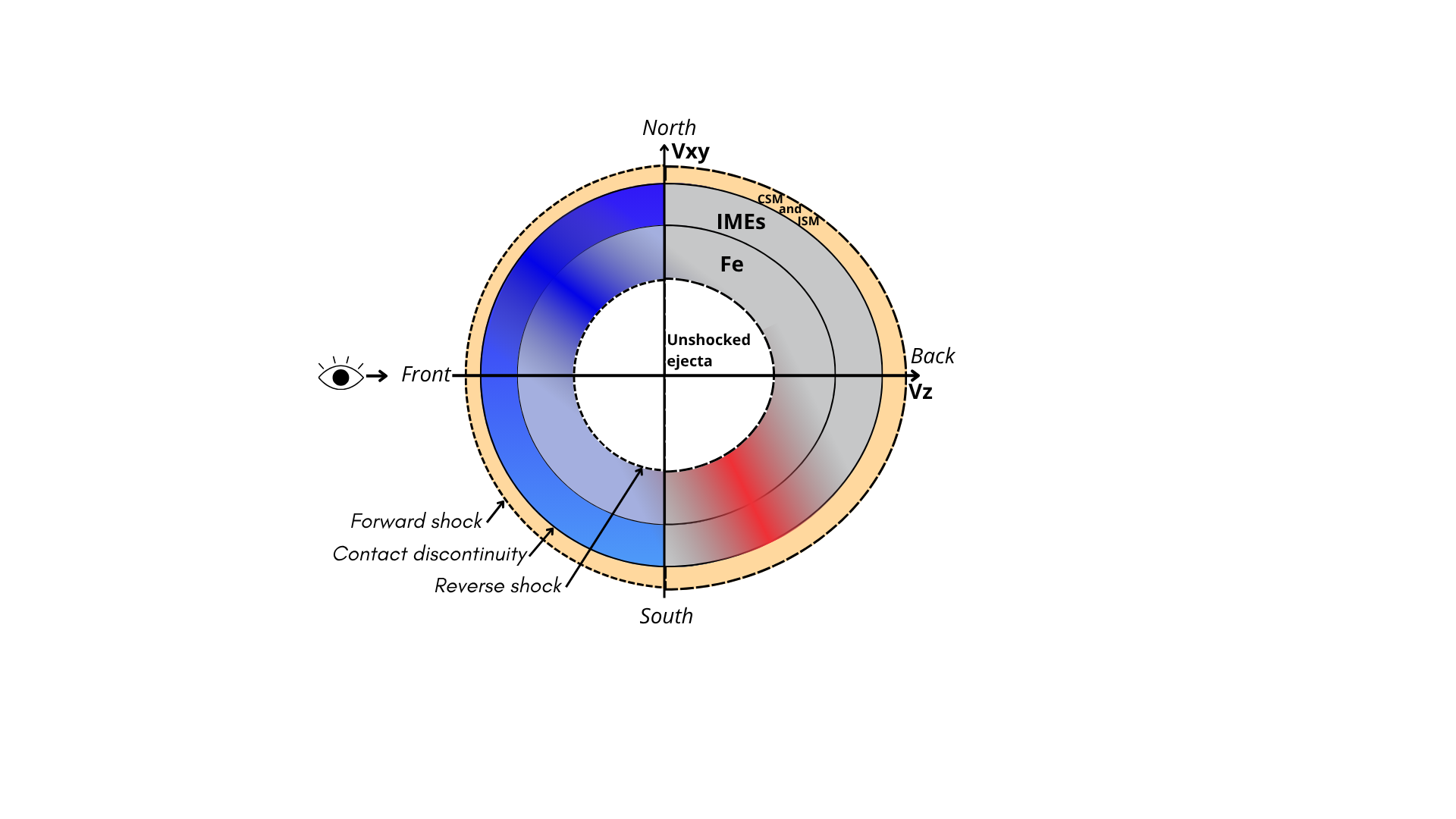}
\caption{\footnotesize Visualization in 3D velocity space summarizing our observations with \emph{XRISM/Resolve} in this study. The dynamics appear asymmetrical, with the back side being faster than the front side. Fe is slower than Si and S. A flux asymmetry is observed: the front side is brighter (shown in blue in the figure) for Si, particularly in the North. This asymmetry is less pronounced for Fe, with more local enhancements (shown in local blue and red in the figure). The gray regions correspond to areas that are fainter along the line of sight.}
\label{fig:Schema4D}
\vspace{-0.2cm}
\end{figure}

The sketch presented in Figure \ref{fig:Schema4D} summarizes the global observations in 3D : the dynamics is asymmetrical, with the back side faster for Si, S and Fe. The Fe dynamics is slower than the IMEs. A strong flux asymmetry between the front and rear side and the north and south, is observed and presented in Figure \ref{fig:CarteRatioNorm}. For Si and S located in the outer layers, the front side appears brighter than the back side : this is represented with the front side in blue and the rear side in gray.
In addition, for Fe, which is more internal, we observe locally structured variations, with the front side sometimes dominant (particularly in the North) and, conversely, the redshifted side being brighter in other regions (particularly in the South). These structures also seem to be present for Si and S in the outer layers, as illustrated in Figure \ref{fig:Schema4D} using blue and red local highlights.

A possible interpretation of this complex 3D morphology and dynamics for Tycho's SNR is a combination of innate and acquired properties. We can address this with 3D numerical simulations of supernova remnants initialized from 3D supernova models based on different explosion mechanisms \citep{Ferrand2019, Ferrand2021, Ferrand2022, Fujimaru2026}.
The more internal flux inhomogeneities, more pronounced for Fe, would originate from the explosion. For example, delayed detonation models can produce inhomogeneous Fe-rich regions in the core, depending on the initial number of ignition points \citep{Seitenzahl2013, Fujimaru2026}. 
The properties of the outer layers, rich in Si, would be more affected by interaction with the surrounding medium. A slight overdensity on the front side might explain why it is brighter and why its velocity is lower. The fact that the velocity stratification between Si, S and Fe is the same in the front and rear sides, is in favor of Tycho's SNR evolving in an asymmetric environment, rather than due to the explosion mechanism. 
However, the asymmetry in velocity and flux in the outer layers may also be explained by innate properties of the progenitor system and explosion model. Some models produce a bulk motion of the remnant as a whole, in which case the remnant is expected to appear brighter in the direction of this motion, even in a homogeneous ambient medium. An alternative explanation for the flux asymmetry between the front and back sides in Si and S is that the initial density distribution of the material was itself asymmetric, as predicted by some models.

\section{Conclusion}

The young Tycho supernova remnant, originating from a Type Ia supernova, was partially covered by five pointings during the first two years of operation of the XRISM observatory. The \emph{Resolve} instrument, with its spectral resolution better than 5 eV, reveals numerous emission lines that are invisible at CCD resolution. The breakthrough for Tycho's SNR lies in the line splitting observed at the center of the remnant: due to the Doppler effect, the front and back sides can now be studied independently.
In this paper, we use a phenomenological model, composed of Gaussian components for each detected line, to obtain a first view of the 3D properties of the remnant at the highest possible spatial resolution using \emph{XRISM/Resolve}. We first apply this model to the global spectra of each pointing, and then at the \emph{Resolve} pixel level, resulting in 149 spectra to fit.
This model allows us to determine the velocity and flux of the front and back sides independently, and this for each element: Si, S, Ar, Ca and Fe. A global line broadening for all the lines of the spectrum is also provided. The new observations with \emph{XRISM/Resolve} open the way to revisit the interpretation of CCD-based observations, in particular regarding the impact of the flux asymmetry on Doppler velocity measurements averaged along the line of sight.

With the \emph{XRISM/Resolve} observations of Tycho's SNR, we reveal that :
\begin{itemize}
\item[$\bullet$] A line-of-sight velocity asymmetry with the back side being faster than the front side for Si, S and Fe.
\item[$\bullet$] A line-of-sight flux asymmetry with the front side being brighter than the back side. For Si and S, the front side can be up to twice as bright as the back side towards the northeastern rim, and decreases towards the southwestern rim. For Fe, the flux ratio varies spatially.
\item[$\bullet$] A line-of-sight velocity stratification of the elements on both the front and rear shells of the shocked ejecta, with lower velocities for Fe than for Si and S. Ar and Ca are more difficult to assess due to line contamination and less statistics.
\end{itemize}

These results suggest that the front side of the remnant is interacting -or has interacted- with a denser surrounding medium than the rear side.
The observed line-of-sight dynamics, asymmetry, and stratification per element provide  crucial constraints to understand the progenitor, explosion mechanism and environment of Tycho's SNR. Combining the \emph{XRISM/Resolve} line-of-sight velocities with previous \emph{Chandra} proper-motion measurements, we reconstruct the 3D dynamics over 65$\%$ of the remnant. Having 3D numerical simulations of supernovae, including the properties of the progenitor system, would allow one to quantify the expected impact of the geometry of the system and environment of normal type Ia supernovae, such as Tycho's SNR.

A forthcoming analysis based on physical models (Non-equilibrium ionization collisional plasma model like \emph{VNEI}) will be necessary to overcome some of the limitations of the present study. Such an analysis will be computationally challenging when using the BXA tool, due to the large number of spectral channels in \emph{XRISM/Resolve} data and the complexity of the physical models to be developed.

\begin{acknowledgements}
LG, AD, RG, JB, FA acknowledge financial support from the Centre national d’études spatiales (CNES), France (ROR: \url{https://ror.org/04h1h0y33}) within the framework of the XRISM and NewAthena missions. C.B. acknowledges support from Chandra Theory grant Nos. TM0-21004X and TM1-22004X, and XRISM Guest Scientist grant No. 80NSSC23K0634. GF acknowledges support from RIKEN-iTHEMS-Report-26.
\end{acknowledgements}

\bibliographystyle{aa} 
\bibliography{Bibliography} 

\begin{appendix} 
\section{Rest Energy}
\label{Appendix : Eref}

\begin{figure}[h!]
\centering
\includegraphics[scale=0.52, trim = 0 0 0 0, clip=true]{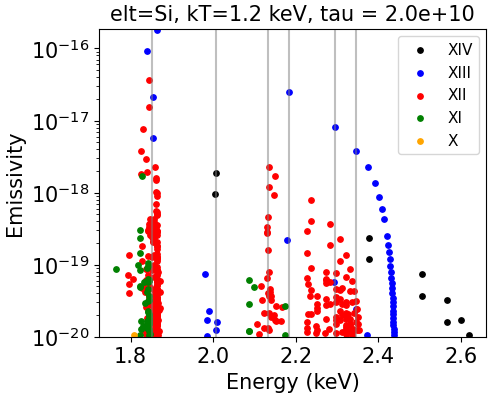}
\includegraphics[scale=0.52, trim = 0 0 0 0, clip=true]{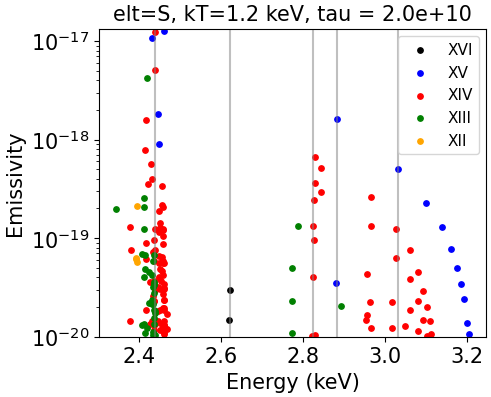}
\caption{\footnotesize 
Emissivity versus energy of individual lines produced by electronic transitions (all the points in these figures) from AtomDB library for Si (\emph{top panel}) and S (\emph{bottom panel}), assuming a temperature of 1.2 keV and an ionization timescale of 2e+10 cm$^{-3}$s. The energies at rest that we used in this article for the Si and S lines are indicated by the vertical gray lines.}
\label{fig:AtomDB}
\end{figure}

To define the reference energies at rest, we compare the positions of the observed lines in the \emph{XRISM/Resolve} spectra with the predictions from AtomDB. As shown in Figure \ref{fig:AtomDB}, most of these lines are in fact complexes composed of numerous individual transitions associated with different ions. The rest energy of these individual lines is fixed, but their relative emissivity depends on the temperature (k$T$) and ionization timescale ($tau$). Therefore, the rest energy of the line complex depends on these thermal properties.

For this study, we adopt as the rest energy of each observed line of Si, S, Ar and Ca, the emissivity-weighted mean value of the underlying line complex. The selected values are listed in Table \ref{tab:ListLines}.
The k$T$ and $tau$ values chosen for the IMEs, respectively 1.2 keV and 2e+10 cm$^{-3}$s in Figure \ref{fig:AtomDB} \citep{Godinaud2024}. 
The energy at rest for the Fe-K line complex (6.427 keV) is based on values from \cite{Yamaguchi2014}. This value is consistent with the average measurements of the temperature and ionization time (respectively $\sim$ 8 keV and $\sim$8e+9 cm$^{-3}$ s) for the Fe-rich ejecta of Tycho's SNR \citep{Godinaud2024}.
It should be noted that the assumption that k$T$ and $tau$ for Ar and Ca are identical to those of Si and S \citep[which dominate the spectrum and therefore drive the measurements of][]{Godinaud2024} is not necessarily valid. Nevertheless, to date there are no independent measurements of the Ar and Ca properties for Tycho’s SNR. As a result for these two elements, the Doppler velocity measurements of the front and back sides are strongly dependent on the assumed rest energies.
It should also be noted that the temperature and ionization conditions may vary across the remnant and between the front and rear sides. Based on \cite{Godinaud2024}, these variations do not appear to be significant enough to affect the conclusions of this study.

To ensure that our results for the Si and S are not affected by thermal effects, we perform a double-check analysis of the Si Ly$\alpha$ line at 2.0061 keV, which is not influenced by such effects, as shown in Figure \ref{fig:AtomDB}. This analysis is presented in Appendix \ref{Appendix : SiLyalpha_fit}.

\section{Double-check with the Si Ly$\alpha$ line }
\label{Appendix : SiLyalpha_fit}

\begin{figure}[t!]
\centering
\includegraphics[scale=0.45, trim = 540 480 965 200, clip=true]{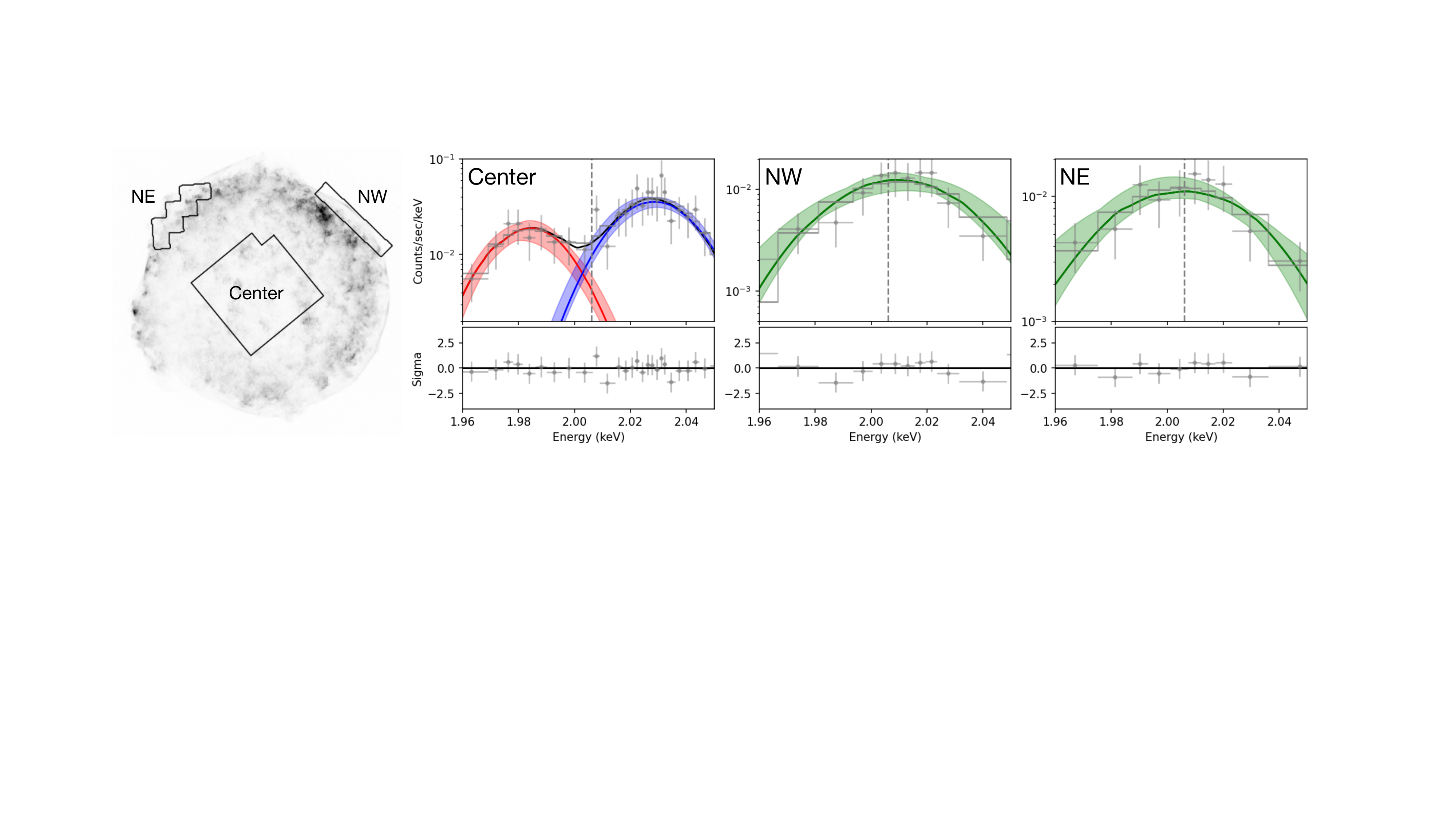}
\caption{\footnotesize Fit of the Si Ly$\alpha$ line between 1.96 and 2.05 keV at the center with two Gaussians.}
\label{fig:IndividualLine}
\end{figure}

As a double-check of the Doppler Si velocity measurements, we restrict the fit to the Si Ly$\alpha$ line between 1.96 and 2.05 keV, since its rest energy is not sensitive to temperature or ionization effects (see Section \ref{Appendix : Eref}). The observed energy centroid of this line depends solely on the ejecta dynamics. However, this line is relatively weak, making it impossible to study at the pixel level. We analyze the center pointing, where the spectrum is fitted with two Gaussian components representing the blueshifted and redshifted emissions, which are clearly separated (see Figure \ref{fig:IndividualLine}). The Doppler shifts and normalizations of these two Gaussians are treated as independent parameters, while their broadenings are linked.

The associated posterior distributions obtained with BXA are very well constrained, as well as the residuals (see Figure \ref{fig:IndividualLine}).
We find that the front and back velocities in the Center are respectively -3245 $\pm$ 215 km/s and 3597 $\pm$ 354 km/s. We therefore observe an asymmetry along the line of sight confirming the results from the global pointing (Section \ref{subsection:Global fit}) and pixel-level studies (Section \ref{subsection:Results-Vz}). The uncertainties are nevertheless larger when using only this line compared to the broadband model used in Section \ref{subsection:Global fit} (see Table \ref{tab:FitGlobal}).
For the central pointing in this global fitting, we note that the normalization ratio shows a significant asymmetry for Si : 62 ($\pm$5) \% of the flux is coming from the front side.
The measured line broadening is 13.6 $\pm$ 1.2 eV.

\section{Normalization radial profiles to detect the model degeneracy near the edges}
\label{Appendix : Norme profile}

\begin{figure}[h!]
\centering
\includegraphics[scale=0.35, trim = 0 0 0 0, clip=true]{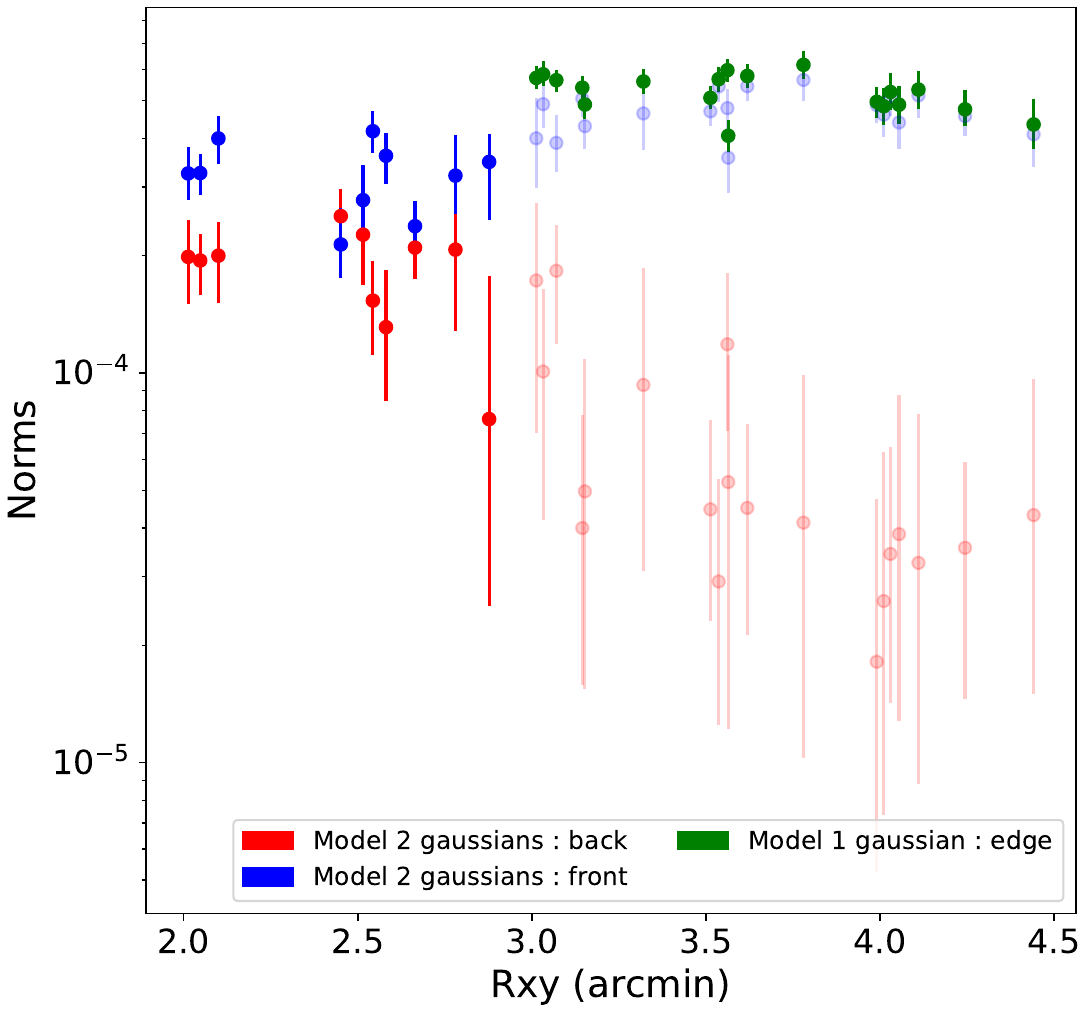}
\caption{\footnotesize Radial profile of the normalization for the NorthWest pointing.
The results obtained with the two-Gaussian model are shown in red and blue, with transparent points where the one-Gaussian (in green) is preferred. 
}
\label{fig:NormProfile}
\end{figure}

Near the edges of the remnant, the lines from the front and back sides do not show a clear separation. The two-Gaussian model can therefore no longer be properly constrained, the line associated with the blueshifted emission systematically becomes dominant, with a normalization consistent with that obtained for the single line in the one-Gaussian model, as illustrated in Fig. \ref{fig:NormProfile}. Conversely, the normalization that should be associated with the back (redshifted) side in the two Gaussians model becomes negligible. The measured redshift is then unconstrained, with uncertainties comparable to the prior used in the fit for this parameter.

We use these observations to define which pixels will use the one-Gaussian model or the two-Gaussian model. To do so, we fit all 149 spectra with the one-Gaussian and the two-Gaussian models. In addition to comparing the C-stat values of the fits and their posterior distributions, we use the radial profiles of the normalizations, as shown in Fig. \ref{fig:NormProfile}. When the results are affected by the degeneracy described previously, we adopt the one-Gaussian model. The final distribution of the models used is shown in Fig. \ref{fig:ModelUsed}.

\section{Example of a posterior distribution}
\label{Appendix : CornerPlot}

\begin{figure*}[h!]
\centering
\includegraphics[scale=0.22, trim = 0 0 0 0, clip=true]{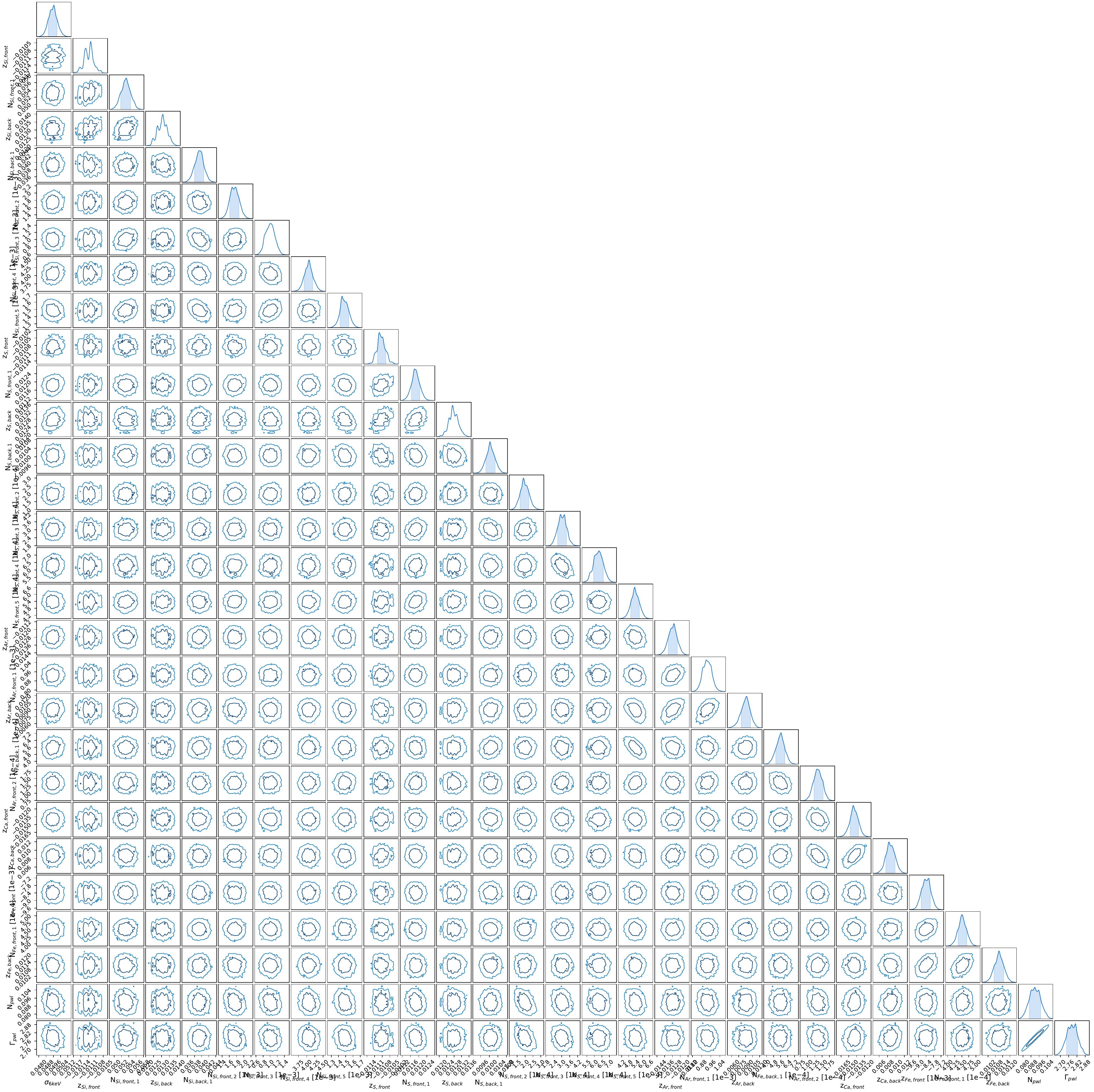}
\caption{\footnotesize Example of a posterior distribution obtained with BXA for the fit of the global Center pointing, combining all its pixels.}
\label{fig:ExCornerPlot}
\end{figure*}

Fig. \ref{fig:ExCornerPlot} shows the posterior distribution for the fit of the spectrum extracted from the global central pointing, whose spectrum is displayed in Fig. \ref{fig:FirstViewSpectra}. This corner plot allows us to check for possible correlations among the 32 parameters of the two-Gaussian model used to fit this spectrum. 

The visualization of posterior distributions is a major strength of BXA, as it enables us to assess the quality of the chosen model and the resulting fit. We have access to such corner plots for all the 149 fits performed in this study, corresponding to each spectrum extracted from the \emph{XRISM/Resolve} pixels.

\section{Ar and Ca velocity radial profiles}
\label{Appendix : Radial profile}

\begin{figure*}[h]
\centering
\includegraphics[scale=0.35, trim = 0 0 0 0, clip=true]{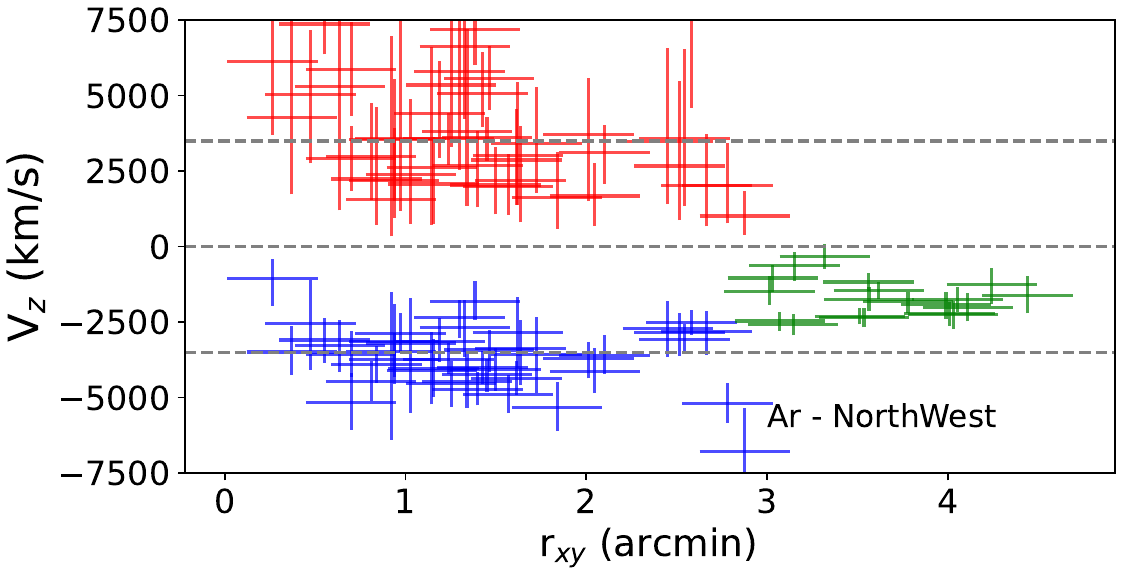}
\includegraphics[scale=0.35, trim = 0 0 0 0, clip=true]{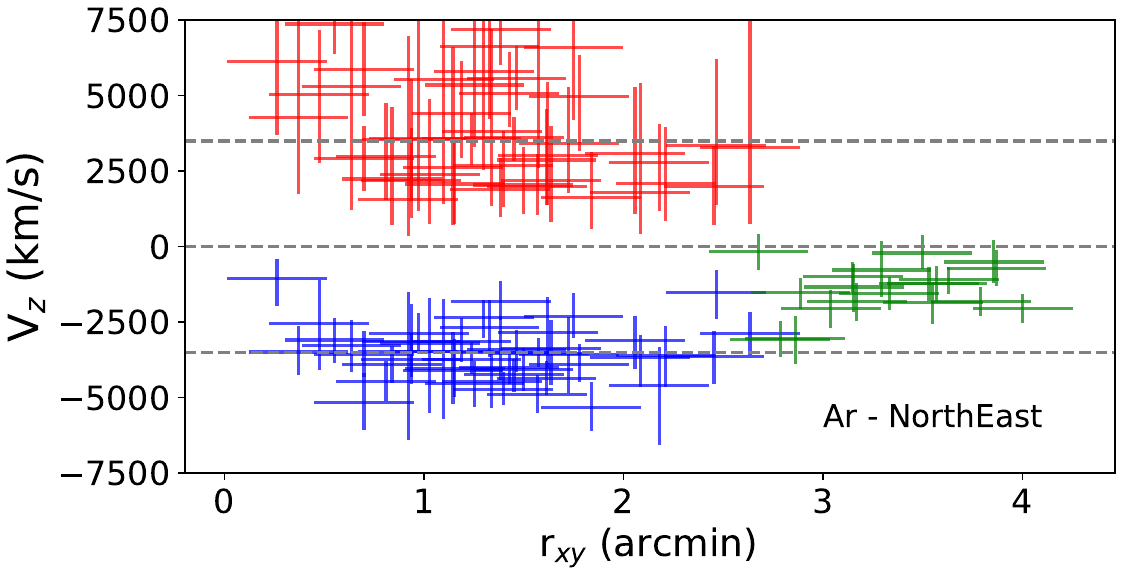}
\includegraphics[scale=0.35, trim = 0 0 0 0, clip=true]{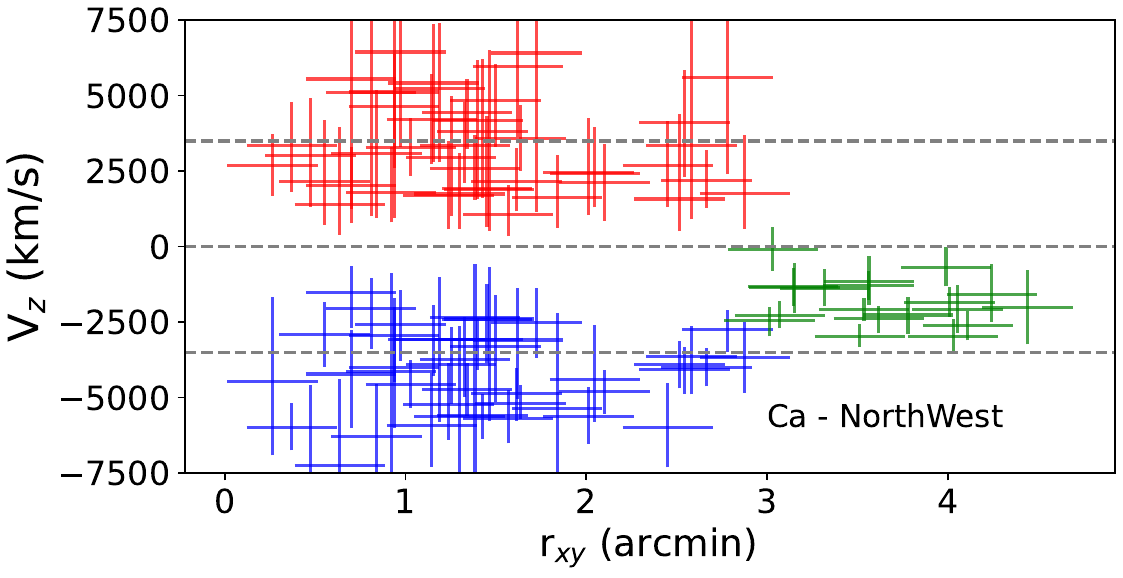}
\includegraphics[scale=0.35, trim = 0 0 0 0, clip=true]{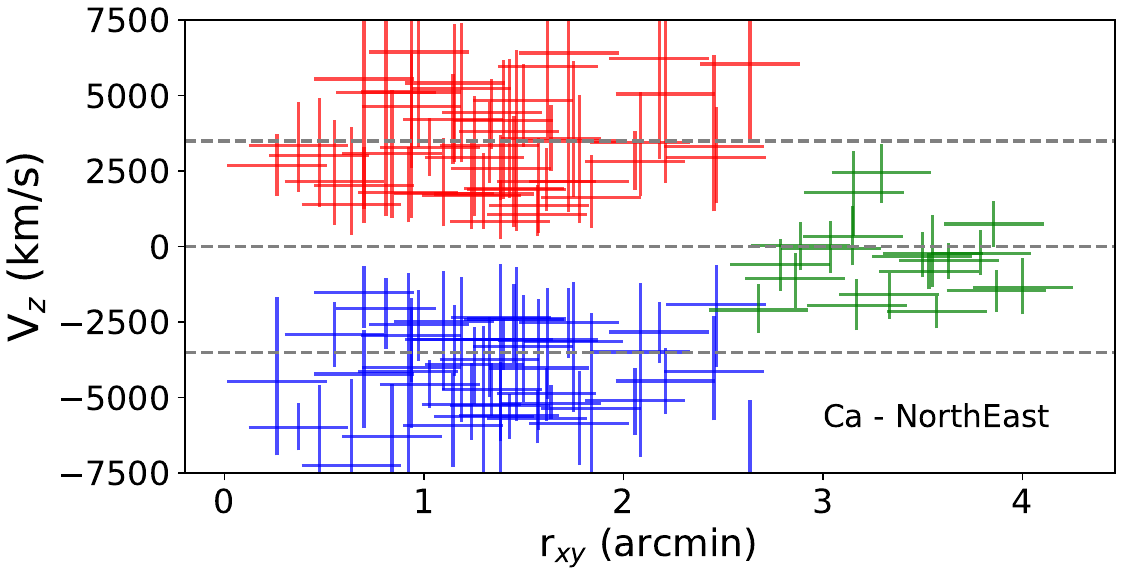}
\caption{\footnotesize 
Radial velocity profiles from the Center toward the NorthWest \emph{(left column)} and toward the NorthEast \emph{(right column)} pointings for the Ar \emph{(first row)} and Ca \emph{(second row)}. Blue and red points correspond to the two-Gaussian model, while green points correspond to the single-Gaussian model.}
\label{fig:Profileradiaux_appendix}
\end{figure*}

Figure \ref{fig:Profileradiaux_appendix} shows the radial velocity profiles for Ar and Ca, for both the NorthWest and NorthEast regions. We find that the dynamics of those elements are difficult to constrain at pixel level, due to statistics and contamination from adjacent lines. In addition, as the energies at rest of these elements are difficult to determine (see Section \ref{Appendix : Eref}), it is unclear whether the shift toward negative velocities at the edges and the apparent velocity asymmetry in the center are real.

\end{appendix}

\end{document}